# MOFClassifier: A Machine Learning Approach for Validating Computation-Ready Metal-Organic Frameworks

Guobin Zhao[1], Pengyu Zhao[1], Yongchul G. Chung[1,2]*

[1]School of Chemical Engineering, [2]Graduate School of Data Science, Pusan National University, Busan 46241, South Korea

**ABSTRACT:** The computational discovery and design of new crystalline materials, particularly metal-organic frameworks (MOFs), heavily relies on high-quality, "computation-ready" structural data. However, recent studies have revealed significant error rates within existing MOF databases, posing a critical data problem that hinders efficient high-throughput computational screening. While rule-based algorithms like MOSAEC, MOFChecker, and the Chen and Manz method (Chen-Manz) have been developed to address this, they often suffer from inherent limitations and misclassification of structures. To overcome this challenge, we developed MOFClassifier, a novel machine learning approach built upon a positive-unlabeled crystal graph convolutional neural network (PU-CGCNN) model. MOFClassifier learns intricate patterns from perfect crystal structures to predict a "crystal-likeness score" (CLscore), effectively classifying MOFs as computation-ready. Our model achieves a ROC value of 0.979 (previous best 0.912) and, importantly, can identify subtle structural and chemical errors that are undetectable by current rule-based methods. By accurately recovering previously misclassified false-negative structures, MOFClassifier reduces the risk of overlooking promising material candidates in large-scale computational screening campaigns. This user-friendly tool is freely available and has been integrated into the preparation workflow for the updated CoRE MOF DB 2025 v1.0, contributing to accelerated computational discovery of MOF materials.

High-fidelity "computation-ready" (CR) crystal structural data are essential for computational screening, evaluation, and design of new crystalline materials.[1] In the field of nanoporous materials, metal-organic frameworks (MOFs) are solid extended framework crystalline materials composed of metal nodes and organic linkers, assembled into various topological structures.[2, 3] Because of the modular nature of the material, MOFs have been investigated for gas storage[4, 5] and separation,[6] catalysis,[7-9] and sensor applications.[10, 11] Their modular nature leads to a combinatorial explosion of possible materials. Consequently, the number of experimentally synthesized MOFs has grown rapidly over the past two decades,[12] driving significant community efforts to computationally screen existing (and hypothetical) MOFs for targeted applications.[13-16] This effort is largely facilitated by open-access databases like the Computation-Ready, Experimental MOF Database (CoRE MOF DB) [17,18]. The structures in the CoRE MOF DBs are mainly derived from the crystallographic information files (CIFs) deposited in the Cambridge Structural Database (CSD).[19] However, direct usage of the CIFs from the CSD poses challenges for computational modeling due to crystallographic disorders and residual solvents. Automated algorithms have been developed to remove solvents prior to prepare databases prior to high-throughput computational screening.[20]

Despite these efforts, many structures in MOF DBs contain chemically invalid structural entries.[21-27] Such "not computation-ready" (NCR) entries significantly hinder high-throughput computational screening, contributing to a critical data problem in computational material science. Consequently, several algorithms have been developed to classify crystal structures as NCR. These includes the Chen-Manz method (bond-order based)[27], MOFChecker (crystal geometry and EQeq charge-based)[22], and more recent MOSAEC (formal charge analysis-based).[28] However, these rule-based algorithms are often incomplete and inherently limited, frequently misclassifying structures (identifying CR as NCR, and vice versa). For instance, well-known MOFs like SBMOF-1[29] and Cu-BTC[30] were mislabeled by MOFChecker and the Chen-Manz method, largely due to their reliance on bond-level analysis and sensitivity to atomic radii values.[20] This highlights the need for a more generalized and robust method to accurately classify computation-ready structures.

To overcome the inherent limitations of rule-based methods, we developed a machine learning approach based on Positive-Unlabeled Crystal Graph Convolutional Neural Networks (PU-CGCNN). This method leverages graph neural networks to capture complex structural relationships and employs a positive-unlabeled learning framework, suited for structural data where only a subset of CR structures is explicitly labeled. Unlike rule-based algorithms that rely on predefined parameters, PU-CGCNN learns implicitly from patterns in large number of structurally perfect frameworks, enabling it to identify subtle chemical and structural errors that are fundamentally undetectable by previous approaches. This learning framework has recently been applied to predict synthesis likelihood of inorganic crystal structures[31,32-34] and MOFs.[35]

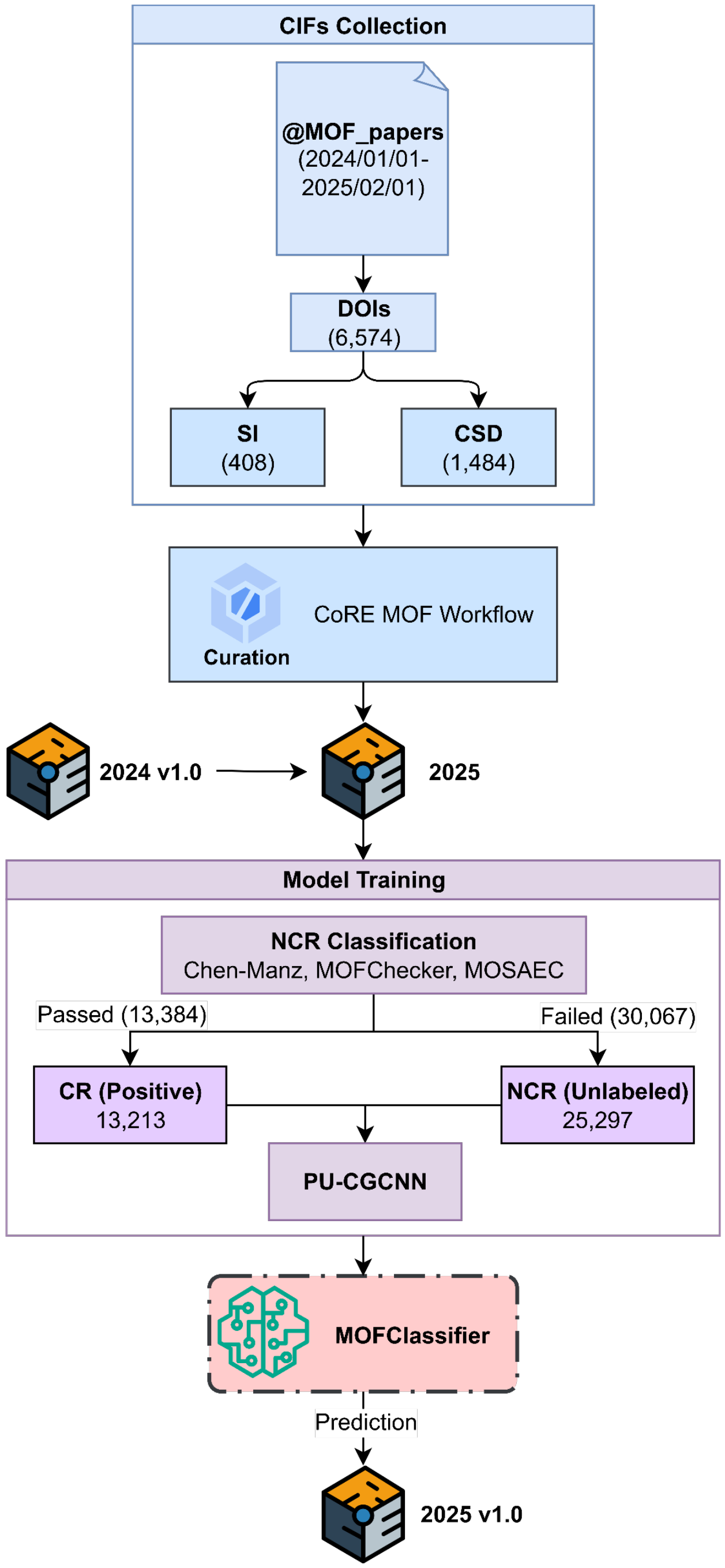

**Figure 1**. Workflow for CoRE MOF DB 2025 v1.0 updates, dataset preparation, and PU-CGCNN model training. Numbers indicate CIF counts, except 6,574 (number of papers).

We applied a PU-CGCNN algorithm to classify MOFs as computation-ready or not. This was achieved by first creating a robust computation-ready dataset using three established error-checking algorithms. The CoRE MOF DB 2025 was prepared by incorporating structures published between January 1st, 2024 and February, 1st, 2025, following the CoRE MOF DB 2024 v1.0 curation steps.[20,36] Subsequently, NCR classification was performed using existing error-checking algorithms to define CR datasets. Details of the error checking tools are provided in the **Supporting Information S2**. Using this dataset, the MOFClassifier model was trained with PU-CGCNN to assign a crystal-likeness score (CLscore) to MOFs.[37-39] The training set comprised 13,213 positive-labeled and 25,297 unlabeled structures. Final scores were assigned based on the average prediction from 100 bootstrap aggregating runs, as shown in **Figure S16**, which demonstrates that convergence was achieved. The overall workflow was illustrated in **Figure 1** and **Figure S15**. Details of the training can be found in **Supporting Information S4**.

As shown in **Figure 2a**, the trained model achieved an AUC of 0.979, outperforming the other three methods. **Figure 2b** presents the recall rate of 0.961 for the positive dataset using a threshold of 0.5, meaning that 12,702 out of 13,213 positive structures were correctly predicted. For structures in the positive dataset with a CLscore (a similarity score with respect to CR structures, ranging from 0 to 1, where values closer to 0 indicate low similarity and values closer to 1 indicate high similarity) below 0.5, most fell within the 0.3–0.5 range, indicating that they are highly similar to CR structures. These 511 structures were further analyzed in detail. In contrast, for the unlabeled dataset, over 50% of the structures exhibit a CLscore below 0.05 (**Figure 2c**), suggesting substantial structural deviation from CR structures or a high degree of imperfection (significant or multiple disorders). Notably, 12.1% of the unlabeled structures were predicted as CR-like, many of which included known “good” structures such as Cu-BTC, which feature geometrically exposed metal sites (open metal sites) or longer bond lengths, leading to false-negative labels using rule-based algorithms. Moreover, the model significantly outperformed random guessing (area under the curve, AUC = 0.5), and the number of data points with a CLscore near 0.5 is relatively small. As shown in **Table S2** and **Figure S17**, varying the cutoff values used to classify MOFs revealed an inverse relationship between the cutoff and recall—recall value dropped noticeably starting from a cutoff of 0.5. The F1 score, which balances recall and precision, reached its maximum at a cutoff of 0.6. Therefore, we considered 0.5 to be a reliable threshold for distinguishing CR from NCR structures. The cutoff can be adjusted depending on the specific objective. For instance, if the objective is to identify error-free structures for computationally demanding high-throughput computational screening, the cutoff can be increased to 0.6–0.7 to ensure higher structural quality of the dataset. Conversely, if the focus is on exploring structure–property relationships, a lower cutoff may be adopted to include a more structurally diverse set of MOFs.

We investigated several representative cases demonstrating MOFClassifier’s capabilities. For instance, in the BAKGIF[40] structure with an incorrect tetrazole protonation state, the MOFClassifier assigned a CLscore of 0.39, which increased to 0.63 upon correction (**Figure 2d**). While MOSAEC also identified this error (**Figure S11**), other errors checking algorithms based on geometry-based or bond-order-based cannot detect such protonation issues. Furthermore, manual inspection of 3.9% of initially positive samples with CLscores < 0.5 revealed that approximately 20% (110) contained crystallographic disorder or chemically invalid frameworks, validating MOFClassifier’s ability to identify subtle issues. A critical "double-negative yields a false-positive" case is MAXJEC_ASR_pacman,[41] where a missing $NH_4^+$ cation and a misidentified hydroxyl group (as $H_2O$ due to crystallographic disorder) lead to overall charge balance. While geometry- or bond-order-based methods were ineffective for charge imbalance (**Supporting Information S2**), MOSAEC also failed to correctly classify this material due to disordered hydrogen atoms resulted in apparent charge balance. The CLscore of the manually fixed MAXJEC structure (MAXJEC_fix_relaxed) increased to 0.93, as shown in **Figure S19**. Crucially, this complex error could currently only be identified by MOFClassifier developed in this work. A similar case is OPUMOE_FSR_pacman (**Figure S21**),[42] involving $CH_3O$ group disorder and ethylaminium ($C_2H_8N^+$) ion loss. We manually classified 110 unlabeled structures as positive and presented the corresponding confusion matrix in **Figure S18**. Based on the CLscore predicted by the MOFClassifier, we identified 401 false negative structures. Additionally, 3,066 positive structures were identified from the unlabeled samples.

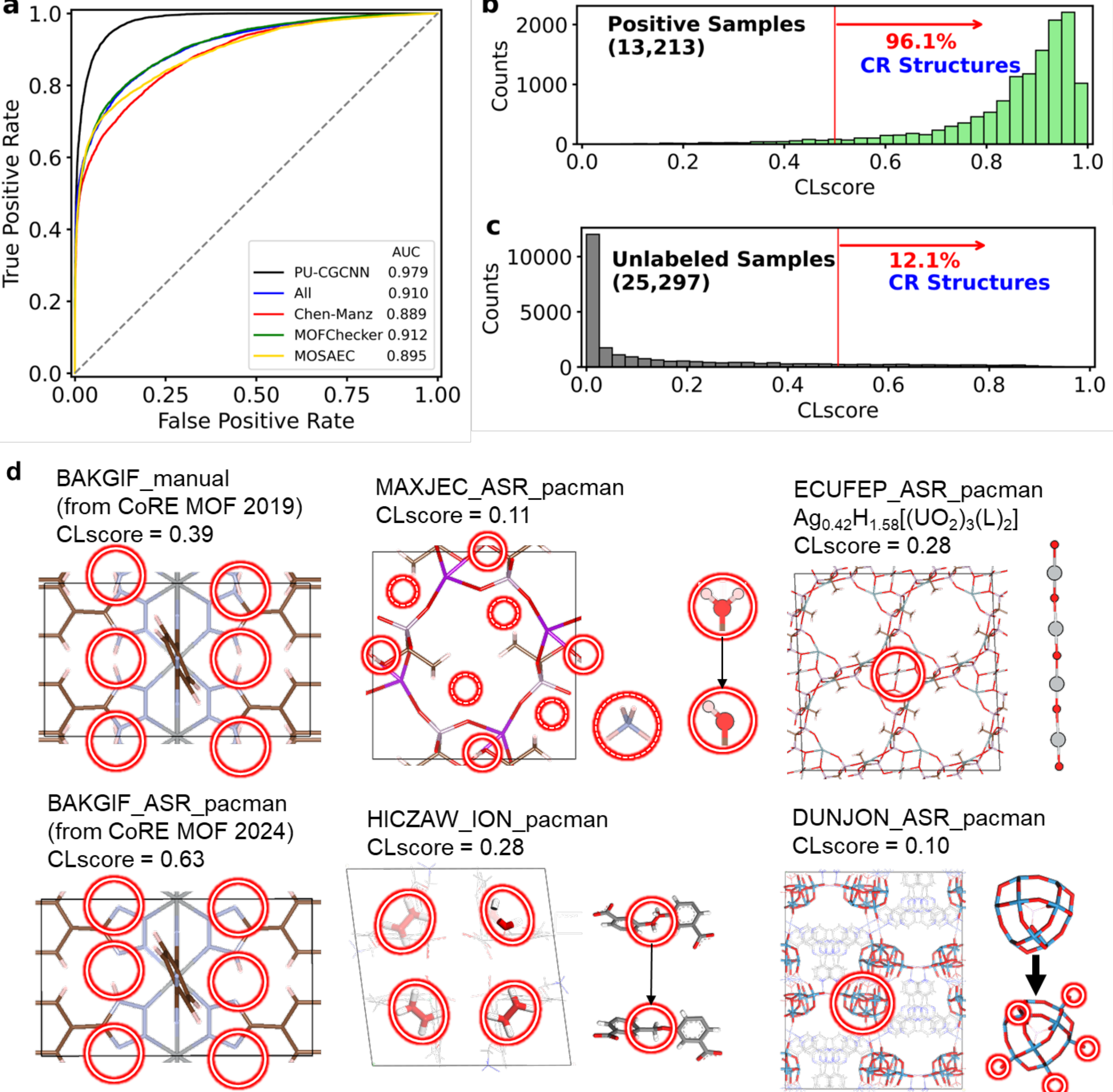

**Figure 2.** (**a**) Receiver operating characteristic (ROC) curves comparing PU-CGCNN model classification performance (CR and NCR) with other methods, including AUC. "All" refers to structures that fail to pass all three individual checkers and are therefore labeled as negative; otherwise, they are labeled as positive. Distribution of predicted crystal structure perfection scores (CLscore) for (**b**) positive samples and (**c**) unlabeled samples. A CLscore threshold of 0.5 (red line) defines high-perfection CR structures. (**d**) Predicted CLscore for BAKGIF (incorrect vs. correct proton counts) and four representative NCR structures from the positive sample set (CLscore <0.5). For BAKGIF, we fixed the number of proton counts from BAKGIF_manual to BAKGIF_ASR_pacman (three BDT units should have 2 protons), more details can be found in CoRE MOF 2024 paper.[20]

Another error that was identified by MOFClassifier is ECUFEP_ASR_pacman (UP-6Ag)[43], which involved incorrect cationic site occupation. Despite the correct overall charge balance, the original CIF for the structure (obtained after coordinated water removal) incorrectly shows only $Ag^+$ ions occupying cationic sites, while the original publication suggests both $Ag^+$ and protons ($H^+$) should be present. This discrepancy highlights MOFClassifier's ability to perform more chemically informed, expert-level assessments beyond simple rule-based checks. MOFClassifier was able to detect subtle ligand-related errors, such as in HICZAW_ION_pacman.[44] Here, a disorder replaced a stable covalent $-CH_2-O-$ linkage with an unstable non-covalent $H_2O-H_2O$ pair. While this structure appeared chemically plausible to rule-based methods (showing no issues in oxidation states, bond orders, or coordination geometry), MOFClassifier successfully identified the error by learning from patterns in correct frameworks. MOFClassifier also identified issues with "retainable terminal oxo groups." For instance, in DUNJON_ASR_pacman,[45] the B-α-$PW_9O_{34}$ cluster is missing terminal oxygen atoms that are neither $H_2O$ nor OH- and should have been retained. MOFClassifier correctly assigned a low CLscore of 0.10 to this structure. Moreover, MOFClassifier also successfully identified apparent disorders (**Figure S20**) and missing counterions required for charge balance (**Figures S21** and **S22**) among positive samples with CLscores below 0.5. For example, in **Figure S22**, the ion was mistakenly removed due to disorder, and MOFClassifier detects this error with CLscore below 0.5. Analysis of false negatives from the three rule-based methods (**Figure S23**) further emphasized MOFClassifier's strength. For instance, Cu-BTC (DOTSOV*)[46] and HAFQUC_FSR_pacman[47], which are structurally correct open metal site (OMS) MOFs, were misclassified but now retained using MOFClassifier.

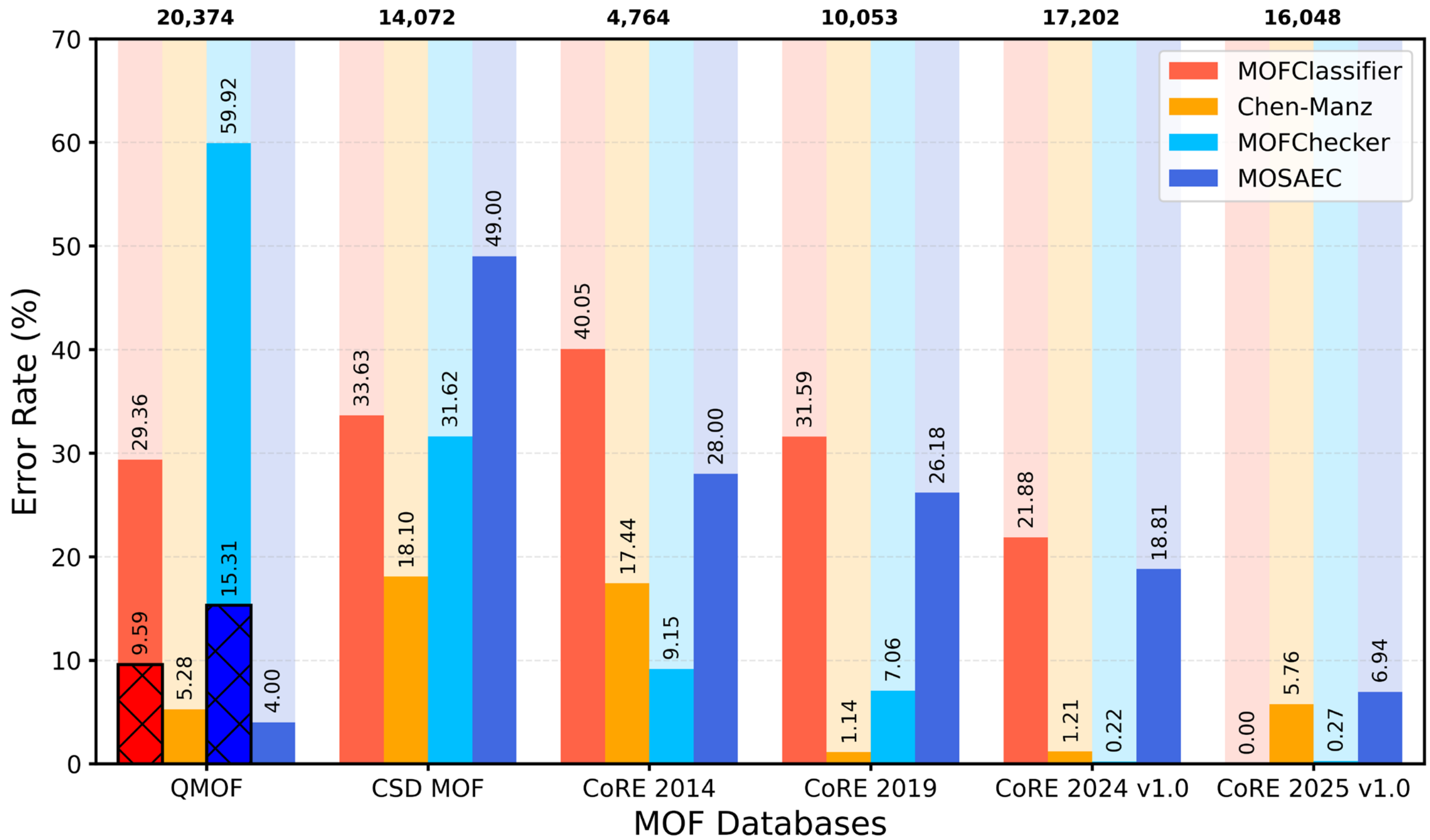

**Figure 3**. Error rates predicted by MOFClassifier and validated by three other methods across experimental, and hybrid MOF databases. MOSAEC results for CoRE MOF 2014, QMOF, and CSD MOF are obtained from published literature (note different CSD MOF Database version v546). CoRE 2019, CoRE 2024 v1.0, and CoRE 2025 v1.0 (checked by MOFClassifier, resulting in an error rate of 0 for MOFClassifier) refer to the CoRE MOF DB series, used as CR datasets in this study. For CoRE 2019 and CoRE 2024 v1.0, we used Chen-Manz and MOFChecker 1.0 (MOFChecker 2.0 used in the figure) to classify the structures. Structures with pore-limiting diameters (PLD) < 2.4 Å were excluded from consideration and are indicated with black shading. The number at the top of the figure represent the number of structures used for classifying. After transferring 1,826 MOF structures from the CoRE MOF 2025 v1.0 CR dataset to the NCR dataset due to failing the Chen-Manz, MOFChecker, and MOSAEC checks (all with an error rate of 0), 14,222 MOFs remain in the CR dataset.

While MOFClassifier provides improved classification over other rule-based methods, it is susceptible to misclassifications, and the rationale for these misclassification is currently not available. For example, we found that SBMOF-1 (OKAYIK_ASR_pacman)[48] was classified as an NCR structure, which may be due to the relatively long Ca–X bond,[49] possibly leading to the Ca atom being identified as isolated. Similar issues arose with alkali and coinage metals (K, Ag, and Li; **Figure S24**). While the Chen-Manz labeled such atoms as "isolated", large atomic radii can also lead to "overbonded" misclassifications. These effects often relegated such structures to unlabeled data, making their rescue as CR challenging without sufficient positive examples.

We benchmarked MOFClassifier across different MOF databases[50, 51] (**Figure 3**). We observe that, as the CoRE MOF DB has been updated and refined, the error rates across all validation methods have significantly decreased. Databases like QMOF[50] contained structures with extremely small or non-existent pores (qmof-d3a3906[52] with 0 Å PLD, **Figure S25**) or even metal salts (qmof-ea0478),[53] with resulted in a CLscore below 0.01. We speculated that this is because MOFClassifier identified the structures as either non-MOFs or non-porous materials. When small-PLD structures were excluded from QMOF, the error rates for MOFClassifier and MOFChecker dropped substantially from 29.36% and 65.21% to 9.59% and 15.31%, respectively. Small PLD values also arose from unremoved guest molecules or solvents, such as ethane and water in qmof-2cf2d31 (SBMOF-1)[48] and qmof-e444654,[54] which block pores. While Chen-Manz and MOFChecker often misclassified these as "isolated", MOSAEC fails to detect such instances as they do not affect metal oxidation states. To address this, we additionally trained a MOFClassifier-QSP model (training with QMOF and CoRE MOF DB, including structures with PLD < 2.4 angstroms) designed to classify structures without considering PLD (**Figure S26**). Additionally, MOFClassifier successfully identified disordered structures that DFT-based methods failed to detect, such as qmof-98c279d (ZJNU-54).[55] We note that the training set consisted entirely of experimentally synthesized structures. Therefore, the current version of MOFClassifier may not be able to detect erroneous structures in as yet synthesized structures. However, such limitations may be addressed by fine-tuning the developed model on hypothetical MOF database using the workflow reported here. Finally, we compared the runtime of the algorithms (**Figure S27** and **Table S4**). We found MOFChecker is the fastest and supports high-throughput prediction, while MOSAEC is the slowest. One caveat is that MOSAEC requires a valid CSD license via its Python API, whereas MOFClassifier is free to use.

In conclusion, the integrity of structural data is key for accelerating the computational discovery and design of new MOF materials. We have developed MOFClassifier, a robust, freely accessible tool based on a PU-CGCNN model, designed to assess the "computation-readiness" of MOF structures by assigning a CLscore. This machine learning approach represents a significant conceptual advance, moving beyond the inherent limitations of rule-based methods by learning complex chemical patterns and identifying subtle structural errors that are previously undetectable. Our evaluation across multiple MOF databases demonstrates that the latest CoRE MOF DB 2025 v1.0, classified with MOFClassifier, exhibits the lowest error rate. This significantly reduces computational resource waste and enhances the reliability of high-throughput virtual screening, thereby accelerating the discovery of high-performing MOFs for experimental synthesis and testing (**Table S5**). We anticipate MOFClassifier will pave the way for similarly intelligent crystal structure validation tools across other complex material systems, accelerating computational design and discovery of novel materials.

## ASSOCIATED CONTENT

**Supporting Information**. Updates to the CoRE MOF DB 2025 v1.0; model parameters for PU-CGCNN, typical examples of results from MOF classification; details of benchmarks for other databases; and additional analyses (PDF). The ZIP file contains: Bagging test results, running time, and flagging information from Chen-Manz, MOFChecker, MOSAEC, and MOFClassifier across different databases; Manual validation results for NCR structures flagged only by MOFClassifier, as well as top-ranked structures; Manually corrected structures. This material is available free of charge via the Internet at http://pubs.acs.org.

## AUTHOR INFORMATION

### Corresponding Author

E-mail: drygchung@gmail.com

### Funding Sources

National Research Foundation of Korea (NRF) grant (RS-2024-00449431). KISTI (KSC-2024-CRE-0283).

### Notes

The authors declare no competing financial interest.
MOFClassifier models and code are available in the following GitHub repository: https://github.com/Chung-Research-Group/MOFClassifier. The script (solvent removal, Chen-Manz, et al) used in this work can be found in https://github.com/mtap-research/CoRE-MOF-Tools.
The CoRE MOF DB 2024 v1.0 and updated CoRE MOF DB (2025 v1.0, 16,036 CR and 27,403 NCR) can be found in Zenodo: https://zenodo.org/records/15055758 and https://zenodo.org/records/15621349. The full data source, prepared files, training results can be found in Zenodo: https://zenodo.org/records/15654431.

## ACKNOWLEDGMENT

This work was supported by the National Research Foundation of Korea (NRF) grant funded by the Korean government (RS-2024-00449431). The authors acknowledge the computational time provided by KISTI (KSC-2024-CRE-0283).

## ABBREVIATIONS

## REFERENCES

(1) Jablonka, K. M.; Ongari, D.; Moosavi, S. M.; Smit, B. Big-Data Science in Porous Materials: Materials Genomics and Machine Learning. *Chemical Reviews* **2020**, *120* (16), 8066-8129. DOI: 10.1021/acs.chemrev.0c00004.

(2) Moghadam, P. Z.; Chung, Y. G.; Snurr, R. Q. Progress toward the computational discovery of new metal-organic framework adsorbents for energy applications. *Nature Energy* **2024**, *9* (2), 121-133. DOI: 10.1038/s41560-023-01417-2.

(3) Kirlikovali, K. O.; Hanna, S. L.; Son, F. A.; Farha, O. K. Back to the Basics: Developing Advanced Metal-Organic Frameworks Using Fundamental Chemistry Concepts. *ACS Nanoscience Au* **2023**, *3* (1), 37-45. DOI: 10.1021/acsnanoscienceau.2c00046.

(4) Chen, Z.; Kirlikovali, K. O.; Idrees, K. B.; Wasson, M. C.; Farha, O. K. Porous materials for hydrogen storage. *Chem-Us* **2022**, *8* (3), 693-716. DOI: 10.1016/j.chempr.2022.01.012.

(5) Yuvaraj, A. R.; Jayarama, A.; Sharma, D.; Nagarkar, S. S.; Duttagupta, S. P.; Pinto, R. Role of metal-organic framework in hydrogen gas storage: A critical review. *International Journal of Hydrogen Energy* **2024**, *59*, 1434-1458. DOI: 10.1016/j.ijhydene.2024.02.060.

(6) Sriram, A.; Choi, S.; Yu, X.; Brabson, L. M.; Das, A.; Ulissi, Z.; Uyttendaele, M.; Medford, A. J.; Sholl, D. S. The Open DAC 2023 Dataset and Challenges for Sorbent Discovery in Direct Air Capture. *ACS Central Science* **2024**, *10* (5), 923-941. DOI: 10.1021/acscentsci.3c01629.

(7) Navalón, S.; Dhakshinamoorthy, A.; Álvaro, M.; Ferrer, B.; García, H. Metal-Organic Frameworks as Photocatalysts for Solar-Driven Overall Water Splitting. *Chemical Reviews* **2023**, *123* (1), 445-490. DOI: 10.1021/acs.chemrev.2c00460.

(8) Pascanu, V.; González Miera, G.; Inge, A. K.; Martín-Matute, B. Metal-Organic Frameworks as Catalysts for Organic Synthesis: A Critical Perspective. *J. Am. Chem. Soc.* **2019**, *141* (18), 7223-7234. DOI: 10.1021/jacs.9b00733.

(9) Jia, T.; Gu, Y.; Li, F. Progress and potential of metal-organic frameworks (MOFs) for gas storage and separation: A review. *Journal of Environmental Chemical Engineering* **2022**, *10* (5), 108300. DOI: 10.1016/j.jece.2022.108300.

(10) Ghosh, S.; Rana, A.; Biswas, S. Metal-Organic Framework-Based Fluorescent Sensors for the Detection of Pharmaceutically Active Compounds. *Chem. Mater.* **2024**, *36* (1), 99-131. DOI: 10.1021/acs.chemmater.3c02459.

(11) Shen, Y.; Tissot, A.; Serre, C. Recent progress on MOF-based optical sensors for VOC sensing. *Chem. Sci.* **2022**, *13* (47), 13978-14007. DOI: 10.1039/D2SC04314A.

(12) Moghadam, P. Z.; Li, A.; Liu, X.-W.; Bueno-Perez, R.; Wang, S.-D.; Wiggin, S. B.; Wood, P. A.; Fairen-Jimenez, D. Targeted classification of metal-organic frameworks in the Cambridge structural database (CSD). *Chem. Sci.* **2020**, *11* (32), 8373-8387. DOI: 10.1039/D0SC01297A.

(13) Colón, Y. J.; Snurr, R. Q. High-throughput computational screening of metal – organic frameworks. *Chemical Society Reviews* **2014**, *43* (16), 5735-5749, 10.1039/C4CS00070F. DOI: 10.1039/C4CS00070F.

(14) Zhao, G.; Chen, Y.; Chung, Y. G. High-Throughput, Multiscale Computational Screening of Metal-Organic Frameworks for Xe/Kr Separation with Machine-Learned Parameters. *Ind. Eng. Chem. Res.* **2023**, *62* (37), 15176-15189. DOI: 10.1021/acs.iecr.3c02211.

(15) Cha, J.; Ga, S.; Lee, S.-j.; Nam, S.; Bae, Y.-S.; Chung, Y. G. Integrated material and process evaluation of metal-organic frameworks database for energy-efficient SF6/N2 separation. *Chem. Eng. J.* **2021**, *426*, 131787. DOI: 10.1016/j.cej.2021.131787.

(16) Kang, M.; Yoon, S.; Ga, S.; Kang, D. W.; Han, S.; Choe, J. H.; Kim, H.; Kim, D. W.; Chung, Y. G.; Hong, C. S. High-Throughput Discovery of $Ni(IN)_2$ for Ethane/Ethylene Separation. *Adv. Sci.* **2021**, *8* (11), 2004940. DOI: 10.1002/advs.202004940.

(17) Chung, Y. G.; Camp, J.; Haranczyk, M.; Sikora, B. J.; Bury, W.; Krungleviciute, V.; Yildirim, T.; Farha, O. K.; Sholl, D. S.; Snurr, R. Q. Computation-ready, experimental metal-organic frameworks: A tool to enable high-throughput screening of nanoporous crystals. *Chem. Mater.* **2014**, *26* (21), 6185-6192. DOI: 10.1021/cm502594j.

(18) Chung, Y. G.; Haldoupis, E.; Bucior, B. J.; Haranczyk, M.; Lee, S.; Zhang, H.; Vogiatzis, K. D.; Milisavljevic, M.; Ling, S.; Camp, J. S. Advances, updates, and analytics for the computation-ready, experimental metal-organic framework database: CoRE MOF 2019. *Journal of Chemical & Engineering Data* **2019**, *64* (12), 5985-5998. DOI: 10.1021/acs.jced.9b00835.

(19) Moghadam, P. Z.; Li, A.; Wiggin, S. B.; Tao, A.; Maloney, A. G. P.; Wood, P. A.; Ward, S. C.; Fairen-Jimenez, D. Development of a Cambridge Structural Database Subset: A Collection of Metal – Organic Frameworks for Past, Present, and Future. *Chem. Mater.* **2017**, *29* (7), 2618-2625. DOI: 10.1021/acs.chemmater.7b00441.

(20) Zhao, G.; Brabson, L. M.; Chheda, S.; Huang, J.; Kim, H.; Liu, K.; Mochida, K.; Pham, T. D.; Terrones, G. G.; Yoon, S. CoRE MOF DB: A curated experimental metal-organic framework database with machine-learned properties for integrated material-process screening. *Matter* **2025**. DOI: 10.1016/j.matt.2025.102140.

(21) Ongari, D.; Talirz, L.; Smit, B. Too Many Materials and Too Many Applications: An Experimental Problem Waiting for a Computational Solution. *ACS Central Science* **2020**, *6* (11), 1890-1900. DOI: 10.1021/acscentsci.0c00988.

(22) Jin, X.; Jablonka, K. M.; Moubarak, E.; Li, Y.; Smit, B. MOFChecker: A Package for Validating and Correcting Metal-Organic Framework (MOF) Structures. *Digital Discovery* **2025**. DOI: 10.1039/D5DD00109A.

(23) Altintas, C.; Avci, G.; Daglar, H.; Azar, A. N. V.; Erucar, I.; Velioglu, S.; Keskin, S. An extensive comparative analysis of two MOF databases: high-throughput screening of computation-ready MOFs for $CH_4$ and $H_2$ adsorption. *J. Mater. Chem. A* **2019**, *7* (16), 9593-9608. DOI: 10.1039/C9TA01378D.

(24) Nurhuda, M.; Perry, C. C.; Addicoat, M. A. Performance of GFN1-xTB for periodic optimization of metal organic frameworks. *Phys. Chem. Chem. Phys.* **2022**, *24* (18), 10906-10914. DOI: 10.1039/D2CP00184E.

(25) Coupry, D. E.; Addicoat, M. A.; Heine, T. Extension of the universal force field for metal-organic frameworks. *J Chem Theory Comput* **2016**, *12* (10), 5215-5225. DOI: 10.1021/acs.jctc.6b00664.

(26) Daglar, H.; Gulbalkan, H. C.; Avci, G.; Aksu, G. O.; Altundal, O. F.; Altintas, C.; Erucar, I.; Keskin, S. Effect of metal-organic framework (MOF) database selection on the assessment of gas storage and separation potentials of MOFs. *Angewandte Chemie International Edition* **2021**, *60* (14), 7828-7837. DOI: 10.1002/anie.202015250.

(27) Chen, T.; Manz, T. A. Identifying misbonded atoms in the 2019 CoRE metal-organic framework database. *RSC Adv.* **2020**, *10* (45), 26944-26951. DOI: 10.1039/D0RA02498H.

(28) White, A. J.; Gibaldi, M.; Burner, J.; Mayo, R. A.; Woo, T. K. High Structural Error Rates in "Computation-Ready" MOF Databases Discovered by Checking Metal Oxidation States. *J. Am. Chem. Soc.* **2025**, *147* (21), 17579-17583. DOI: 10.1021/jacs.5c04914.

(29) Banerjee, D.; Simon, C. M.; Plonka, A. M.; Motkuri, R. K.; Liu, J.; Chen, X.; Smit, B.; Parise, J. B.; Haranczyk, M.; Thallapally, P. K. Metal-organic framework with optimally selective xenon adsorption and separation. *Nat. Commun.* **2016**, *7* (1), ncomms11831. DOI: 10.1038/ncomms11831.

(30) Wu, Y.; Kobayashi, A.; Halder, G. J.; Peterson, V. K.; Chapman, K. W.; Lock, N.; Southon, P. D.; Kepert, C. J. Negative Thermal Expansion in the Metal-Organic Framework Material $Cu_3$(1,3,5-benzenetricarboxylate)$_2$. *Angewandte Chemie International Edition* **2008**, *47* (46), 8929-8932. DOI: 10.1002/anie.200803925.

(31) Mordelet, F.; Vert, J.-P. A bagging SVM to learn from positive and unlabeled examples. *Pattern Recognition Letters* **2014**, *37*, 201-209. DOI: 10.1016/j.patrec.2013.06.010.

(32) Jang, J.; Gu, G. H.; Noh, J.; Kim, J.; Jung, Y. Structure-based synthesizability prediction of crystals using partially supervised learning. *Journal of the American Chemical Society* **2020**, *142* (44), 18836-18843.

(33) Gleaves, D.; Fu, N.; Siriwardane, E. M. D.; Zhao, Y.; Hu, J. Materials synthesizability and stability prediction using a semi-supervised teacher-student dual neural network. *Digital Discovery* **2023**, *2* (2), 377-391. DOI: 10.1039/D2DD00098A.

(34) Kim, S.; Schrier, J.; Jung, Y. Explainable Synthesizability Prediction of Inorganic Crystal Polymorphs Using Large Language Models. *Angewandte Chemie International Edition* **2025**, *64* (19), e202423950. DOI: 10.1002/anie.202423950.

(35) Kim, W.-T.; Lee, W.-G.; An, H.-E.; Furukawa, H.; Jeong, W.; Kim, S.-C.; Long, J. R.; Jeong, S.; Lee, J.-H. Machine learning-assisted design of metal – organic frameworks for hydrogen storage: A high-throughput screening and experimental approach. *Chem. Eng. J.* **2025**, *507*, 160766. DOI: 10.1016/j.cej.2025.160766.

(36) Coudert, F.-X. Using Social Media Bots to Keep up with a Vibrant Research Field: The Example of @ MOF_papers. *Chem. Mater.* **2023**, *35* (7), 2657-2660. DOI: 10.1021/acs.chemmater.3c00383.

(37) Xie, T.; Grossman, J. C. Crystal graph convolutional neural networks for an accurate and interpretable prediction of material properties. *Physical review letters* **2018**, *120* (14), 145301. DOI: 10.1103/PhysRevLett.120.145301.

(38) Chen, Y.; Zhao, G.; Yoon, S.; Habibi, P.; Hong, C. S.; Li, S.; Moultos, O. A.; Dey, P.; Vlugt, T. J. H.; Chung, Y. G. Computational Exploration of Adsorption-Based Hydrogen Storage in Mg-Alkoxide Functionalized Covalent-Organic Frameworks (COFs): Force-Field and Machine Learning Models. *ACS Appl. Mater. Interfaces* **2024**, *16* (45), 61995-62009. DOI: 10.1021/acsami.4c11953.

(39) Zhao, G.; Chung, Y. G. PACMAN: A Robust Partial Atomic Charge Predicter for Nanoporous Materials Based on Crystal Graph Convolution Networks. *J Chem Theory Comput* **2024**, *20* (12), 5368-5380. DOI: 10.1021/acs.jctc.4c00434.

(40) Ouellette, W.; Darling, K.; Prosvirin, A.; Whitenack, K.; Dunbar, K. R.; Zubieta, J. Syntheses, structural characterization and properties of transition metal complexes of 5,5'-(1,4-phenylene) bis (1H-tetrazole)($H_2$bdt), 5',5''-(1,1'-biphenyl)-4,4'-diylbis (1H-tetrazole)($H_2$dbdt) and 5,5',5''-(1,3,5-phenylene) tris (1H-tetrazole)($H_3$btt). *Dalton Trans.* **2011**, *40* (45), 12288-12300. DOI: 10.1039/C1DT11590A.

(41) Dong, D.-P.; Liu, L.; Sun, Z.-G.; Jiao, C.-Q.; Liu, Z.-M.; Li, C.; Zhu, Y.-Y.; Chen, K.; Wang, C.-L. Synthesis, Crystal Structures, and Luminescence and Magnetic Properties of 3D Chiral and Achiral Lanthanide Diphosphonates Containing Left- and Right-Handed Helical Chains. *Cryst. Growth Des.* **2011**, *11* (12), 5346-5354. DOI: 10.1021/cg2009368.

(42) Kumar Maka, V.; Tamuly, P.; Jindal, S.; Narasimha Moorthy, J. Control of In-MOF topologies and tuning of porosity through ligand structure, functionality and interpenetration: Selective cationic dye exchange. *Applied Materials Today* **2020**, *19*, 100613. DOI: 10.1016/j.apmt.2020.100613.

(43) Yang, W.; Wu, H.-Y.; Wang, R.-X.; Pan, Q.-J.; Sun, Z.-M.; Zhang, H. From 1D Chain to 3D Framework Uranyl Diphosphonates: Syntheses, Crystal Structures, and Selective Ion Exchange. *Inorg. Chem.* **2012**, *51* (21), 11458-11465. DOI: 10.1021/ic301183h.

(44) Ma, M.-L.; Ji, C.; Zang, S.-Q. Syntheses, structures, tunable emission and white light emitting $Eu^{3+}$ and $Tb^{3+}$ doped lanthanide metal-organic framework materials. *Dalton Trans.* **2013**, *42* (29), 10579-10586. DOI: 10.1039/C3DT50315A.

(45) Song, B.-Q.; Wang, X.-L.; Sun, C.-Y.; Zhang, Y.-T.; Wu, X.-S.; Yang, L.; Shao, K.-Z.; Zhao, L.; Su, Z.-M. An organic – inorganic hybrid photocatalyst based on sandwich-type tetra-Co-substituted phosphotungstates with high visible light photocatalytic activity. *Dalton Trans.* **2015**, *44* (31), 13818-13822. DOI: 10.1039/C5DT01560J.

(46) Wu, Y.; Kobayashi, A.; Halder, G. J.; Peterson, V. K.; Chapman, K. W.; Lock, N.; Southon, P. D.; Kepert, C. J. Negative Thermal Expansion in the Metal – Organic Framework Material Cu3(1,3,5-benzenetricarboxylate)2. *Angewandte Chemie International Edition* **2008**, *47* (46), 8929-8932. DOI: 10.1002/anie.200803925.

(47) Potts, S. V.; Barbour, L. J.; Haynes, D. A.; Rawson, J. M.; Lloyd, G. O. Inclusion of Thiazyl Radicals in Porous Crystalline Materials. *J. Am. Chem. Soc.* **2011**, *133* (33), 12948-12951. DOI: 10.1021/ja204548d.

(48) Plonka, A. M.; Chen, X.; Wang, H.; Krishna, R.; Dong, X.; Banerjee, D.; Woerner, W. R.; Han, Y.; Li, J.; Parise, J. B. Light Hydrocarbon Adsorption Mechanisms in Two Calcium-Based Microporous Metal Organic Frameworks. *Chem. Mater.* **2016**, *28* (6), 1636-1646. DOI: 10.1021/acs.chemmater.5b03792.

(49) Chen, T.; Manz, T. A. A collection of forcefield precursors for metal-organic frameworks. *RSC Adv.* **2019**, *9* (63), 36492-36507. DOI: 10.1039/C9RA07327B.

(50) Rosen, A. S.; Iyer, S. M.; Ray, D.; Yao, Z.; Aspuru-Guzik, A.; Gagliardi, L.; Notestein, J. M.; Snurr, R. Q. Machine learning the quantum-chemical properties of metal-organic frameworks for accelerated materials discovery. *Matter* **2021**, *4* (5), 1578-1597. DOI: 10.1016/j.matt.2021.02.015.

(51) Li, A.; Perez, R. B.; Wiggin, S.; Ward, S. C.; Wood, P. A.; Fairen-Jimenez, D. The launch of a freely accessible MOF CIF collection from the CSD. *Matter* **2021**, *4* (4), 1105-1106. DOI: 10.1016/j.matt.2021.03.006.

(52) Chen, H.-Y.; Zhang, T.-L.; Zhang, J.-G.; Yang, L.; Qiao, X.-J. Synthesis, Characterization and Properties of Tri-substitute Potassium Salt of Trinitrophloroglucinol. *Chinese Journal of Chemistry* **2007**, *25* (1), 59-62. DOI: 10.1002/cjoc.200790017.

(53) Fischer, N.; Klapötke, T. M.; Peters, K.; Rusan, M.; Stierstorfer, J. Alkaline Earth Metal Salts of 5,5'-Bistetrazole-from Academical Interest to Practical Application. *Zeitschrift für anorganische und allgemeine Chemie* **2011**, *637* (12), 1693-1701. DOI: 10.1002/zaac.201100263.

(54) Qian, J.; Hu, J.; Zhang, J.; Yoshikawa, H.; Awaga, K.; Zhang, C. Solvent-Induced Assembly of Octacyanometalates-Based Coordination Polymers with Unique afm1 Topology and Magnetic Properties. *Cryst. Growth Des.* **2013**, *13* (12), 5211-5219. DOI: 10.1021/cg400909b.

(55) Jiao, J.; Dou, L.; Liu, H.; Chen, F.; Bai, D.; Feng, Y.; Xiong, S.; Chen, D.-L.; He, Y. An aminopyrimidine-functionalized cage-based metal – organic framework exhibiting highly selective adsorption of $C_2H_2$ and $CO_2$ over $CH_4$. *Dalton Trans.* **2016**, *45* (34), 13373-13382. DOI: 10.1039/C6DT02150F.

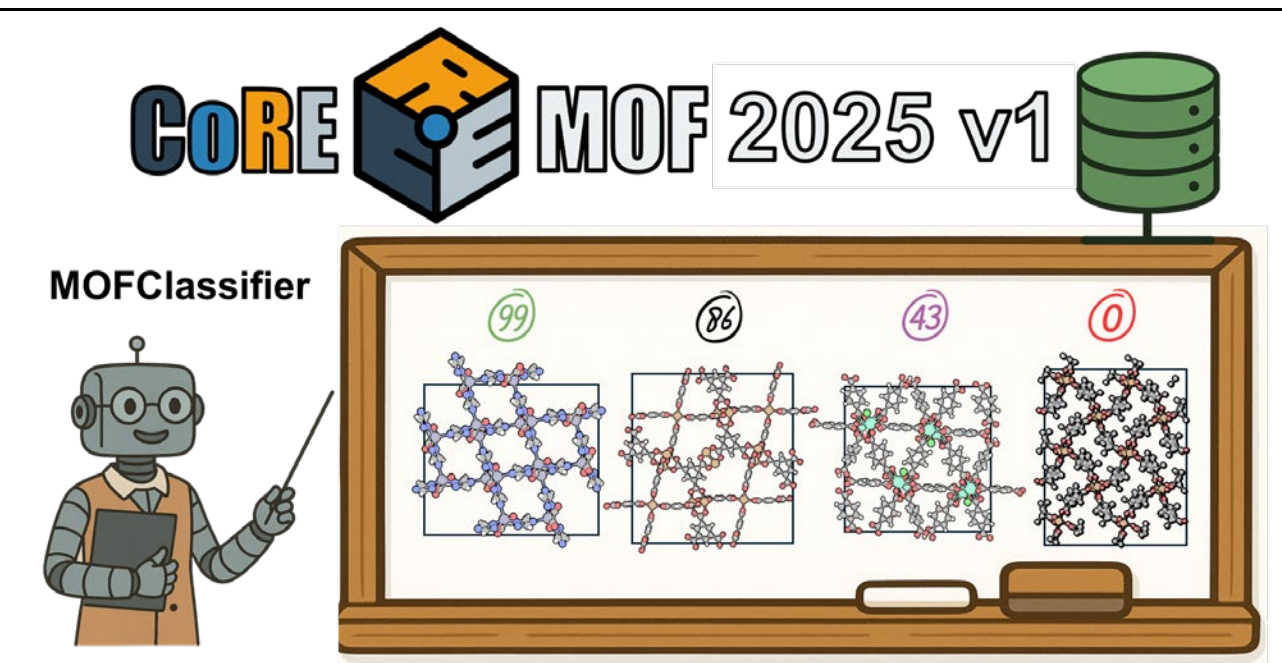

# Supporting Information

## MOFClassifier: A Machine Learning Approach for Validating Computation-Ready Metal-Organic Frameworks

Guobin Zhao[1], Pengyu Zhao[1], Yongchul G. Chung[1,2]*
[1]School of Chemical Engineering, and [2]Graduate School of Data Science, Pusan National University, Busan 46241, South Korea

# S1 CoRE MOF DB Update

Based on previous studies, we found that nearly 99% of the CIFs in the supporting information (SI) were downloaded from the American Chemical Society (ACS) and the Royal Society of Chemistry (RSC). Therefore, in this work, we collected new CIFs from the Cambridge Structural Database (CSD) and SI (ACS and RSC) reported between January 1, 2024, and February 1, 2025, according to @MOF_papers (on X or Bluesky).[1] The source code is available at https://github.com/fxcoudert/PapersBot. The keyword set includes “MOF”, “COF”, “ZIF”, “metal-organic framework”, “covalent-organic framework”, “imidazolate framework”, “porous coordination polymer”, and “framework material”. Note that variations involving hyphens (both “-“ and “–“) and periods (“.”) are also considered. The corresponding DOIs are documented in https://github.com/fxcoudert/MOF_papers/posted.dat. Finally, there are 6,988 ASR, 5,950 FSR, and 446 ION MOFs in the CR dataset updated in this work as shown in **Figure 1**.

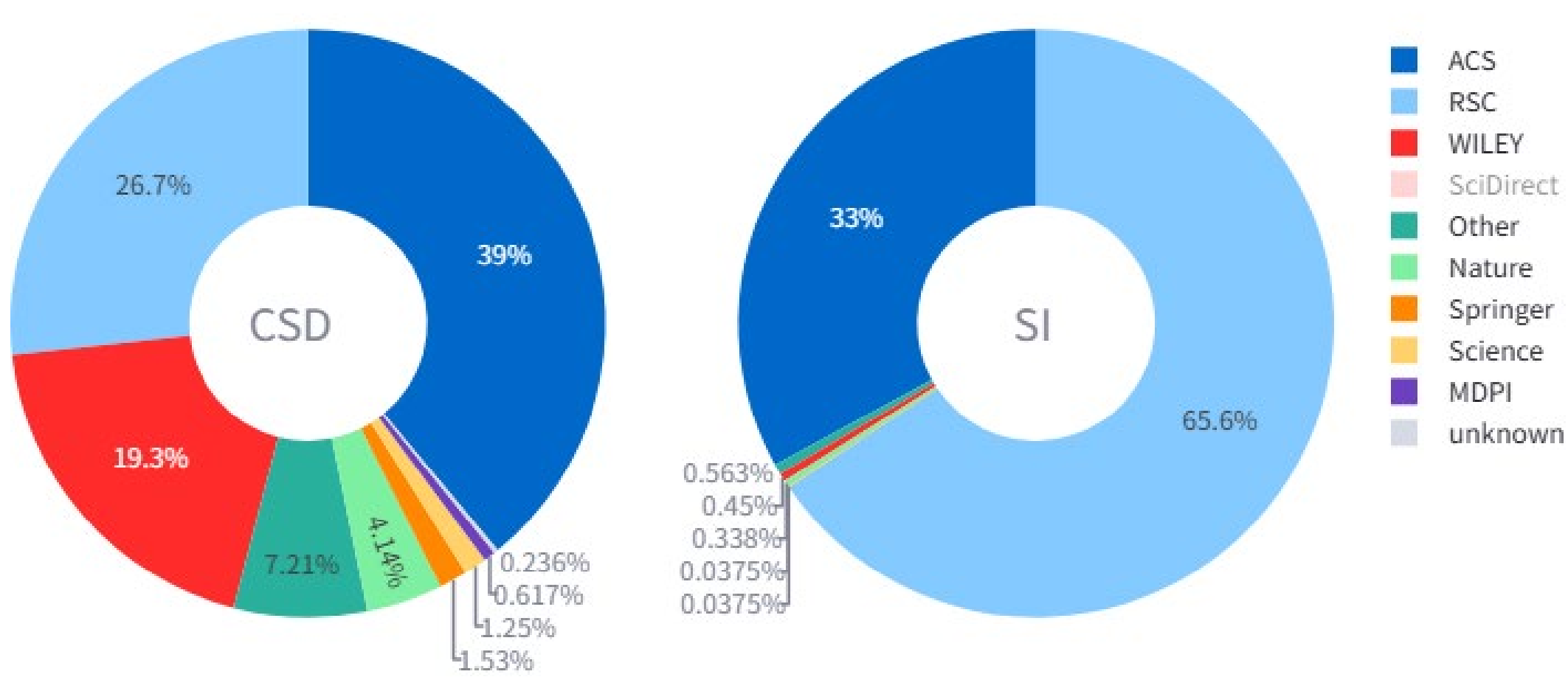

**Figure S1.** The percentage of downloaded sources (CSD: Cambridge Structure Database; SI: supporting information) of MOFs in the CoRE MOF DB.

## S2 Checking Tools

**Table S1**. The check list of three error checking tools.

| Tool | Check List | Bool |
| --- | --- | --- |
| **Chen-Manz**[2] | Over bonded carbon | True |
| | Under bonded carbon | True |
| | Atom overlapping | True |
| | Isolated (not for ION MOF) | True |
| **MOFChecker**[3] | has_atomic_overlaps | True |
| | has_overcoordinated_c | True |
| | has_overcoordinated_n | True |
| | has_overcoordinated_h | True |
| | has_suspicicious_terminal_oxo | True |
| | has_undercoordinated_c | True |
| | has_undercoordinated_n | True |
| | has_undercoordinated_rare_earth | True |
| | has_undercoordinated_alkali_alkaline<br>(only for CoRE MOF DB v8) | True |
| | has_geometrically_exposed_metal | True |
| | has_lone_molecule (not for ION MOF) | True |
| | has_carbon<br>is_porous (for benchmark) | False |
| **MOSAEC**[4] | Impossible | True |
| | Unknown | True |
| | Zero_Valent | True |
| | noint | True |
| | low_prob_1 | True |
| | low_prob_2 | True |
| | low_prob_3 | True |
| | low_prob_multi | True |

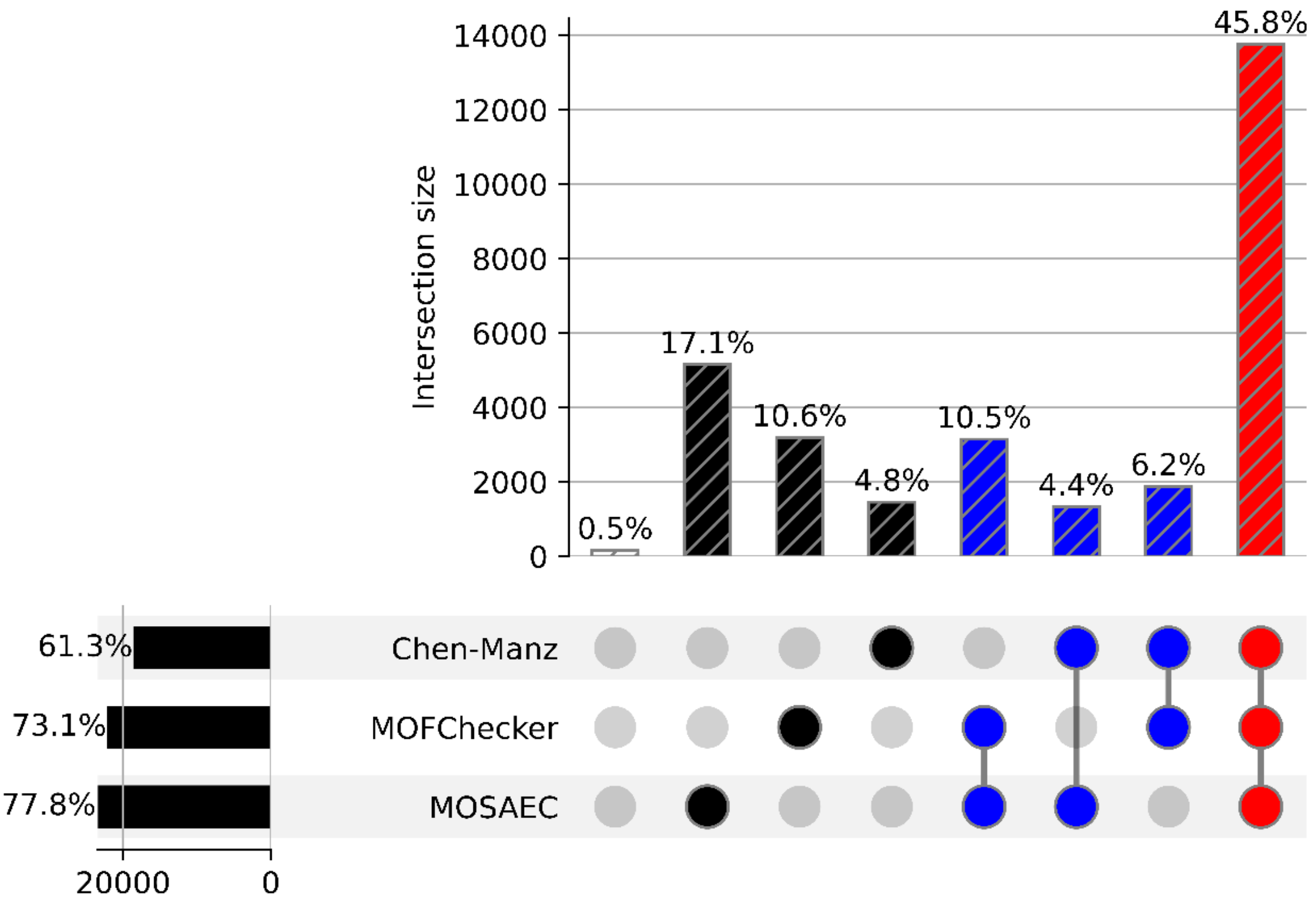

**Figure S2**. Result of NCR structures for CoRE MOF DB checked by "Chen-Manz", MOFChecker, and MOSAEC. There are 0.5% structures defined as NCR due to occupancy of atoms. The total number of NCR structures is 30,067 from CoRE MOF DB 2025 (not classified by MOFClassifier).

## S2.1 "Chen-Manz" Method

We analyzed the NCR structures uniquely identified by each tool to evaluate their respective advantages and disadvantages, as well as to investigate true negative and false negative cases and their underlying causes. For each case, we randomly selected several structures from each error type and determined whether they were false negatives or true negatives. We also ensured, as much as possible, that each structure exhibited only that specific error and was identified as erroneous by only one method. The same applies to **S2.2** and **S2.3**.

For the bond-order-based method (Chen-Manz), the most frequently identified NCR structures were attributed to "over bonded carbon" and "under bonded carbon" (**Figure S3**). Cases involving "isolated" atoms were rare, as all structures had undergone solvent removal using a solvent removal algorithm. Notably, only 4.8% of all NCR structures were uniquely identified by the Chen-Manz method (**Figure S2**).

To further explore this method, we examined representative false negative cases for each NCR type (**Figure S4**):

1. SBMOF-1 was misclassified as containing an isolated atom (Ca). This error arose because the bond distance between Ca and its coordinating atoms was longer than the threshold defined in the algorithm, leading to a failure in bond detection and incorrect assignment as a free atom.

2. In MIL-53, an O-H bond was erroneously flagged due to its bond length. According to the atom type radii (ATR) defined by the "Chen-Manz" method (refer to the original publication and the CoRE MOF 2024 database for all parameters and algorithmic details),[5] the overlap cutoff between the O and H atoms is calculated as 0.635 Å (0.38: ATR for H atom; 0.89: ATR for O atom), which exceeds the distance of O-H in MIL-53 (0.562 Å). As a result, the structure was incorrectly classified as "atom overlapping".

3. PCN-60 features a linker containing a C≡C triple bond. The bond-order calculation did not consider for this triple bond and classified the carbon atoms involved as "under bonded".

4. In NOTT-116, the aromatic ring within the linker was non-planar, with two non-adjacent carbon atoms located in unusually close proximity. This led to a spurious bond assignment between atoms that are not bonded, ultimately causing the structure to be flagged as containing "under bonded carbon".

Several true negative cases highlight the unique advantages of the "Chen-Manz" method. However, upon inspecting all cases classified as "isolated", we found that these were in fact misassignments caused by excessive interatomic distances, rather than genuine chemical isolation. Representative true negative cases (**Figure S5**) from other NCR types are summarized below:

1. In MOF-155-J, an extraneous oxygen atom was incorrectly present within the metal cluster, leading to "over bonded" due to excessive coordination.

2. In MIL-53, a hydrogen atom was missing from the bridging hydroxyl group between metal centers, resulting in an under-bonded configuration within the metal cluster.

3. In UiO-66-$NH_3^+$[$ReO_4^-$], the original structure contained a disordered $ReO_4^-$ anion, which was unintentionally removed by the solvent-removal algorithm. The resulting structure lacked the necessary anion for charge compensation, leading this method to classify the structure as containing "under bonded".

4. ECUT-111 exhibited inverted conformers, in which nitrogen atoms from the two conformers were positioned unusually close to one another. Their interatomic distance fell below the overlap cutoff threshold of 0.86 Å for N-N pairs, resulting in defining as NCR due to perceived atomic overlap.

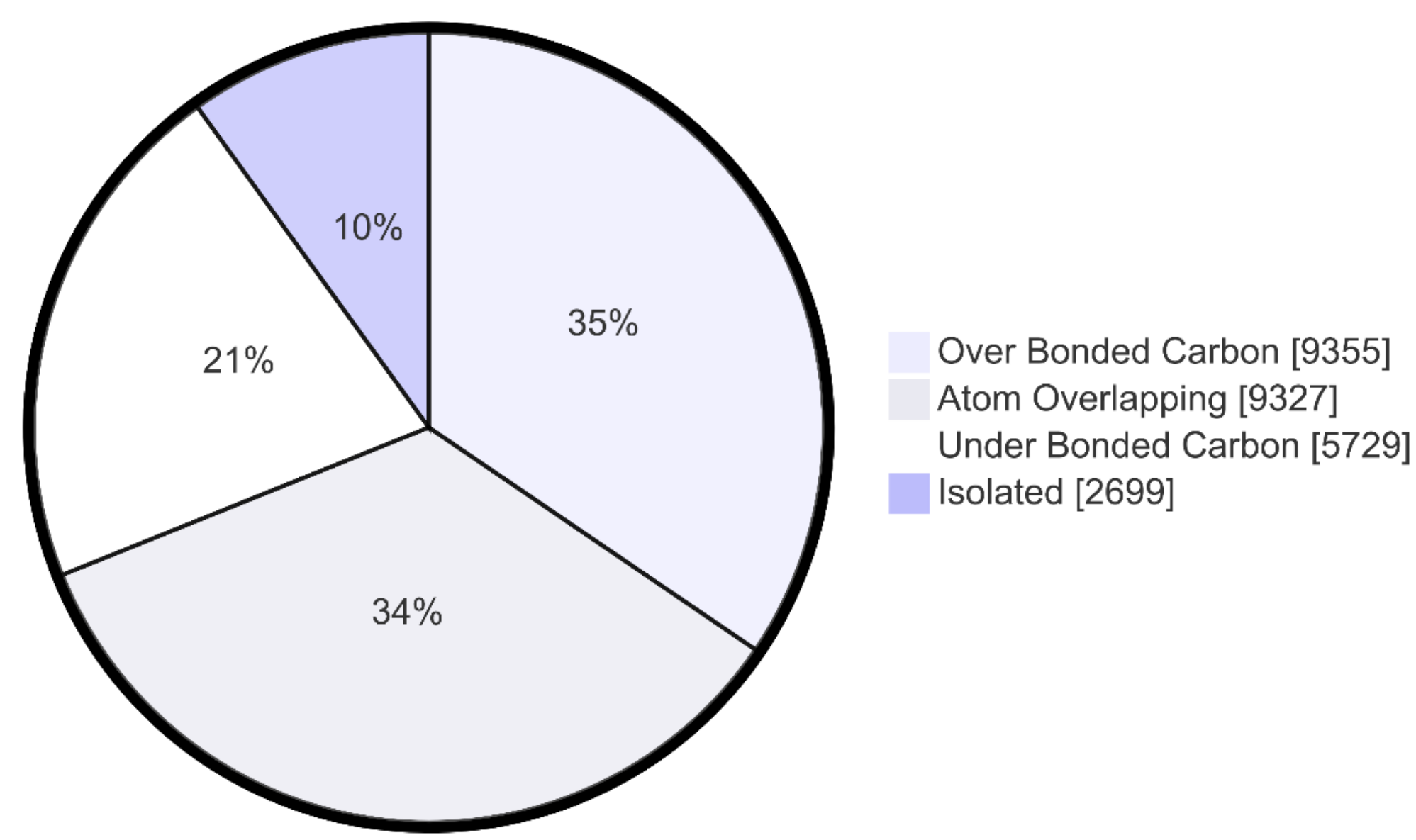

**Figure S3**. Results of each NCR case for CoRE MOF DB defined by "Chen-Manz".

41467_2016_BFncomms11831_MOESM1567_ESM_ASR_pacman (SBMOF-1: “**Isolated**”)

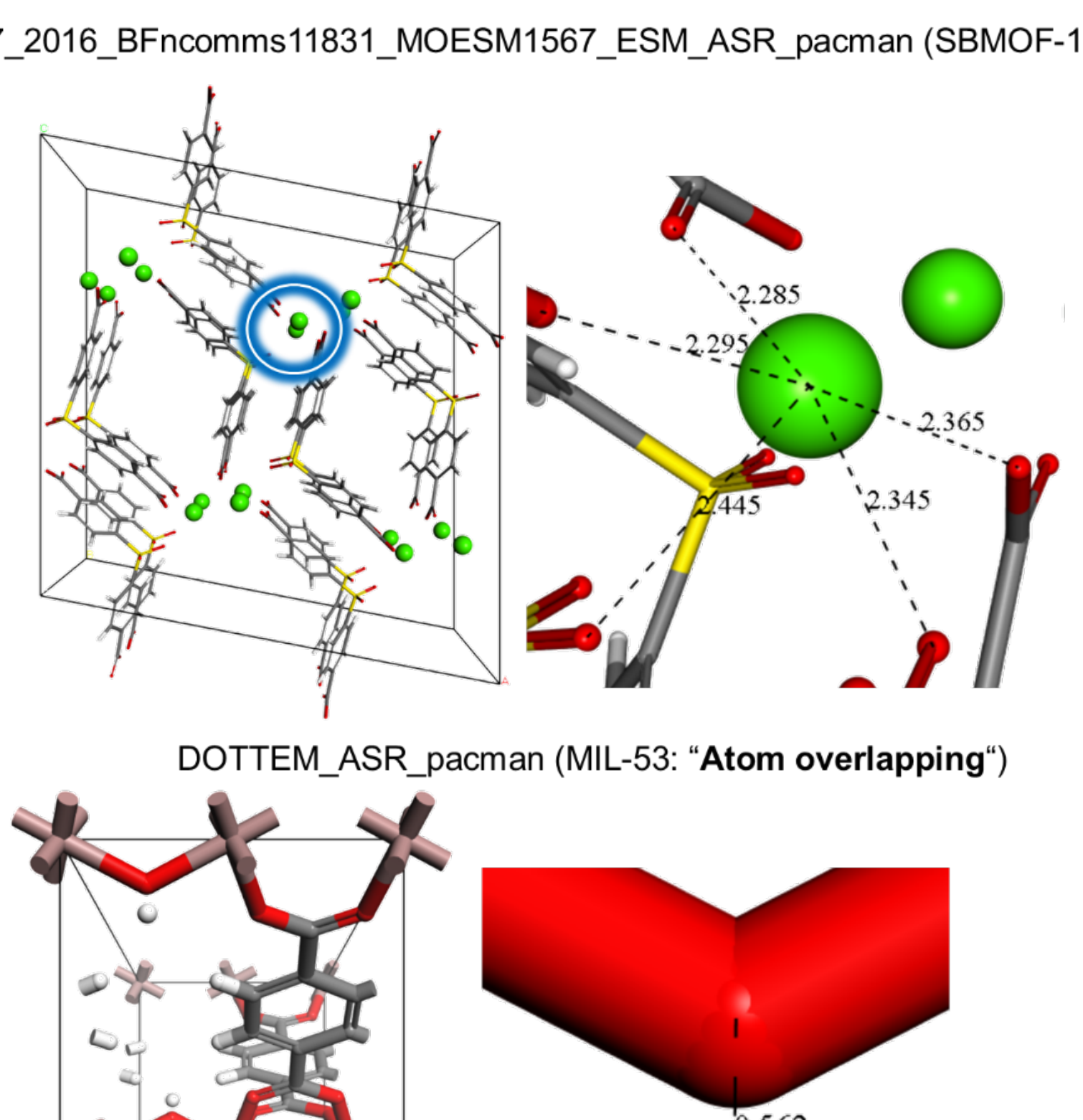

DOTTEM_ASR_pacman (MIL-53: “**Atom overlapping**“)

VUJBEI_ASR_pacman (PCN-60: “**Under bonded carbon**”)

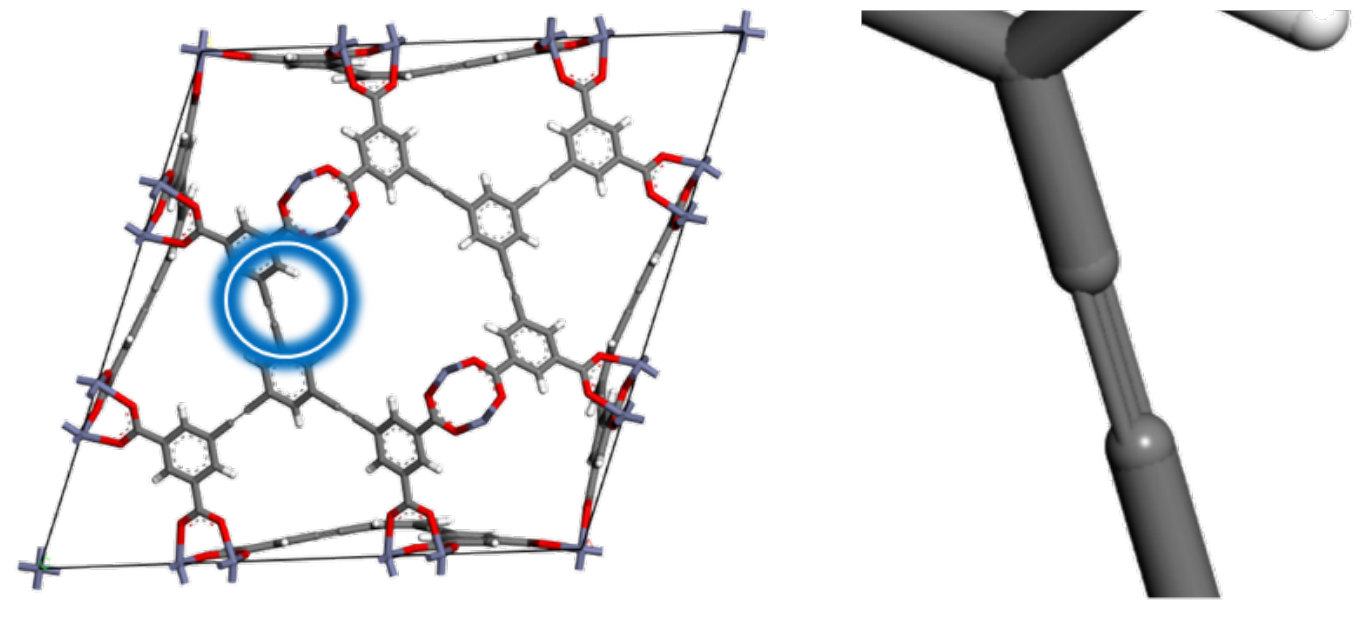

LURRIA_ASR_pacman (NOTT-116: “**Over bonded carbon**”)

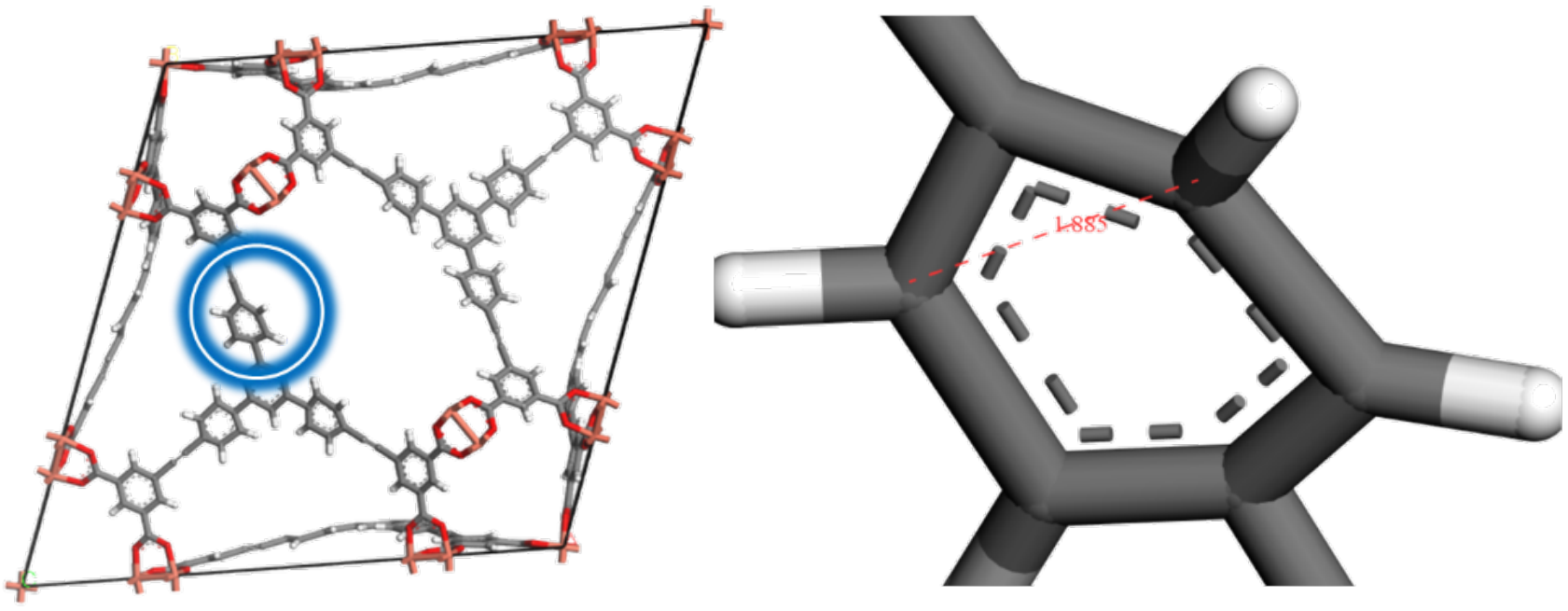

**Figure S4**. Representative NCR structures are misidentified by the “Chen-Manz”.

SUGTEV_ASR_pacman (MOF-155-J: “**Over bonded carbon**”)

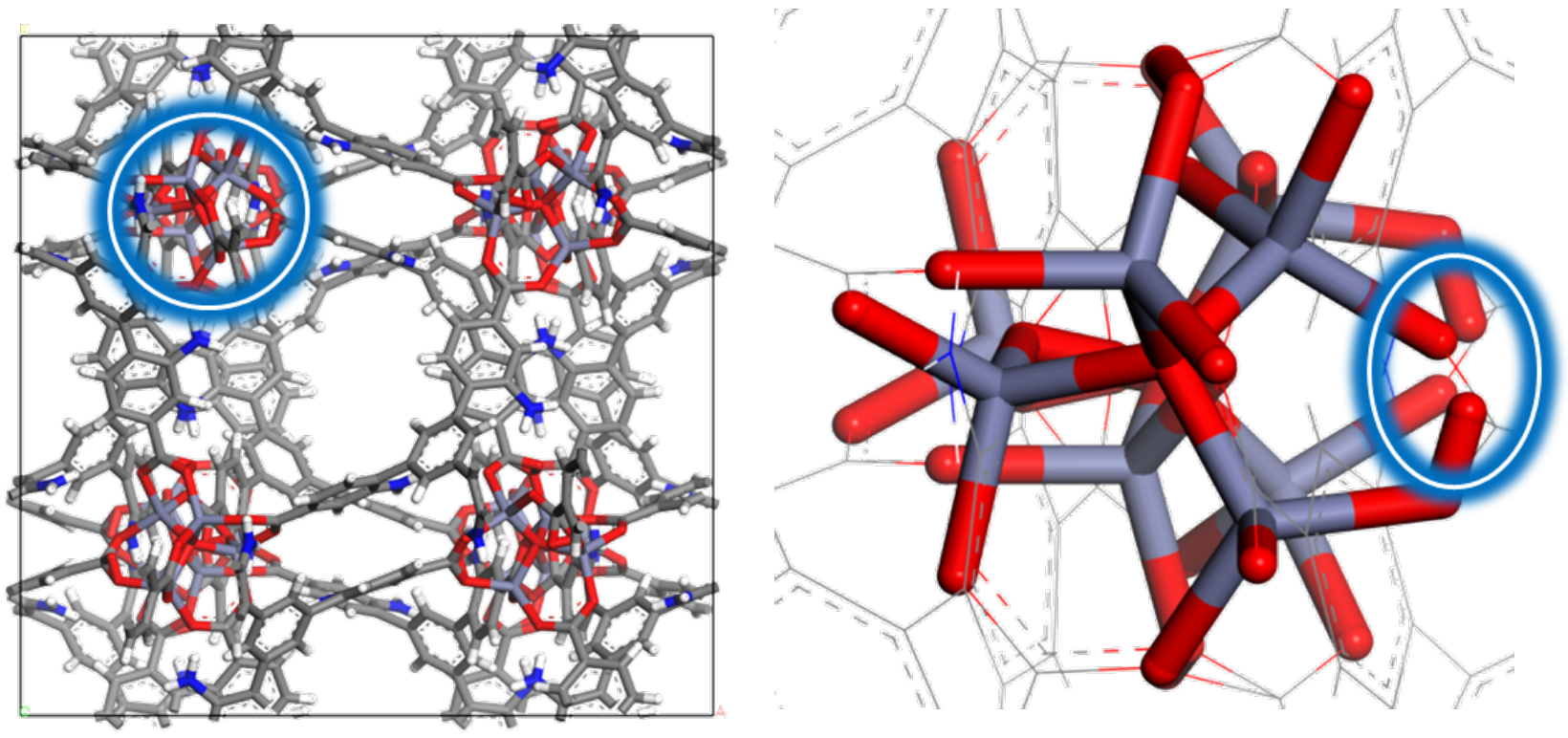

LURLOB_ASR_pacman (MIL-53 deuterium: “**Under bonded carbon**”)

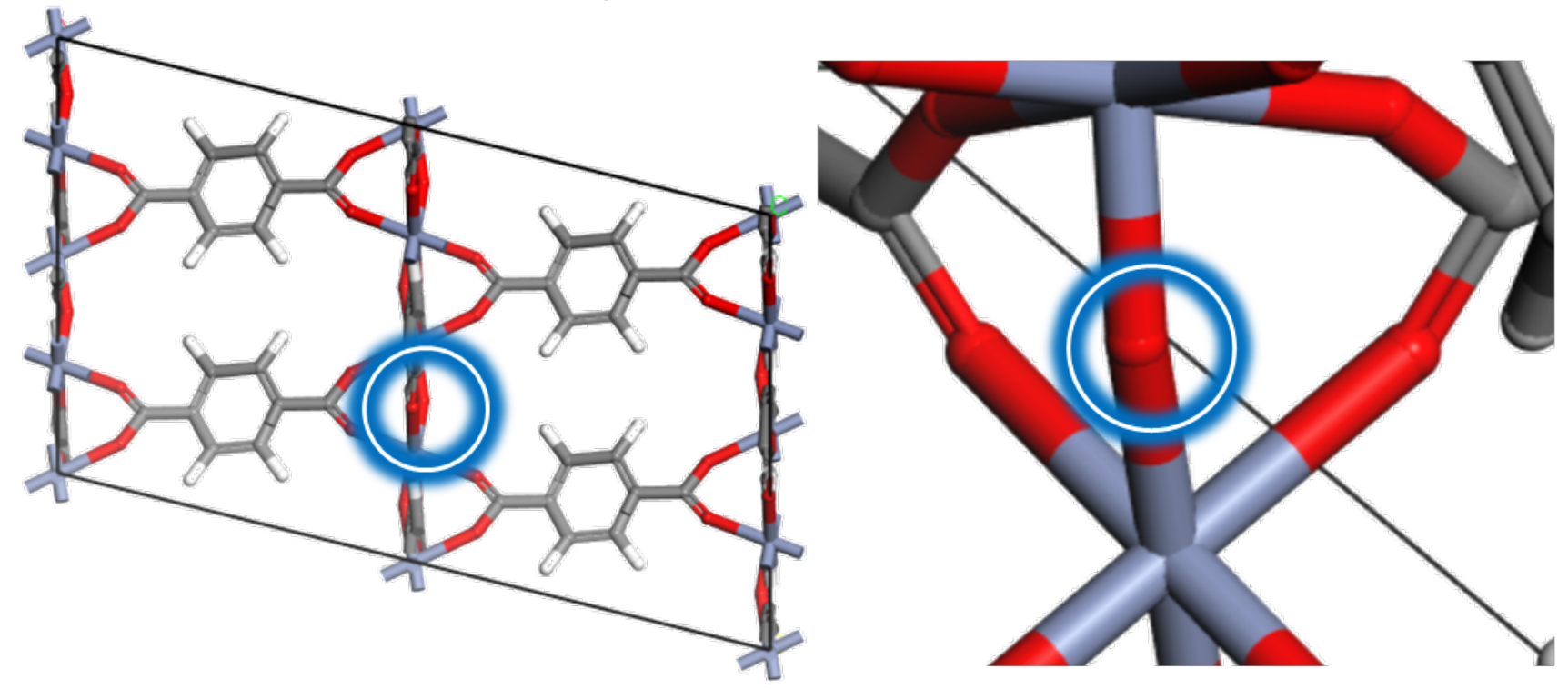

EWILON_FSR_pacman (UiO-66-$NH_3^+ReO_4$: “**Under bonded carbon**”)

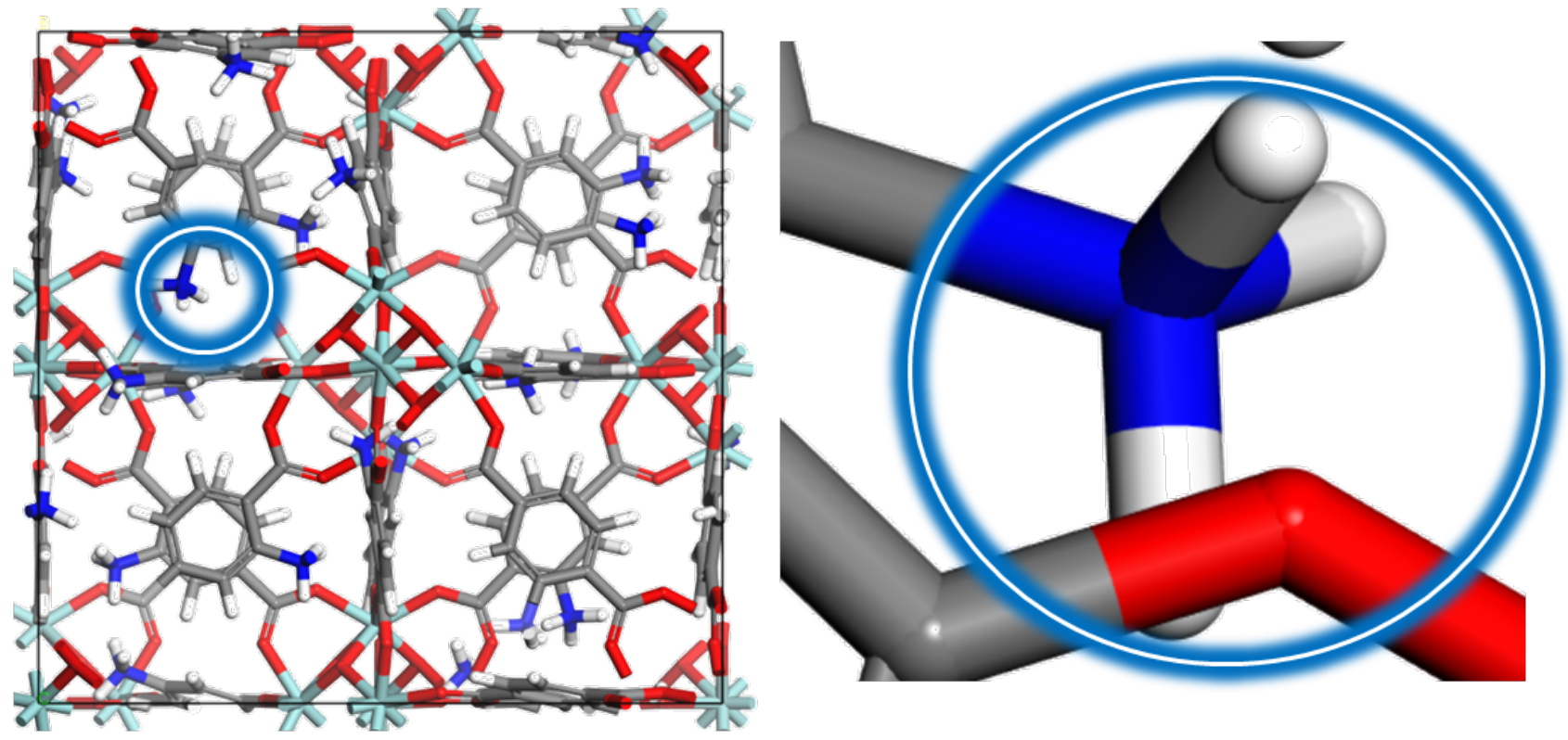

ALUQEG_ASR_pacman (ECUT-111: “**Atom overlapping**“)

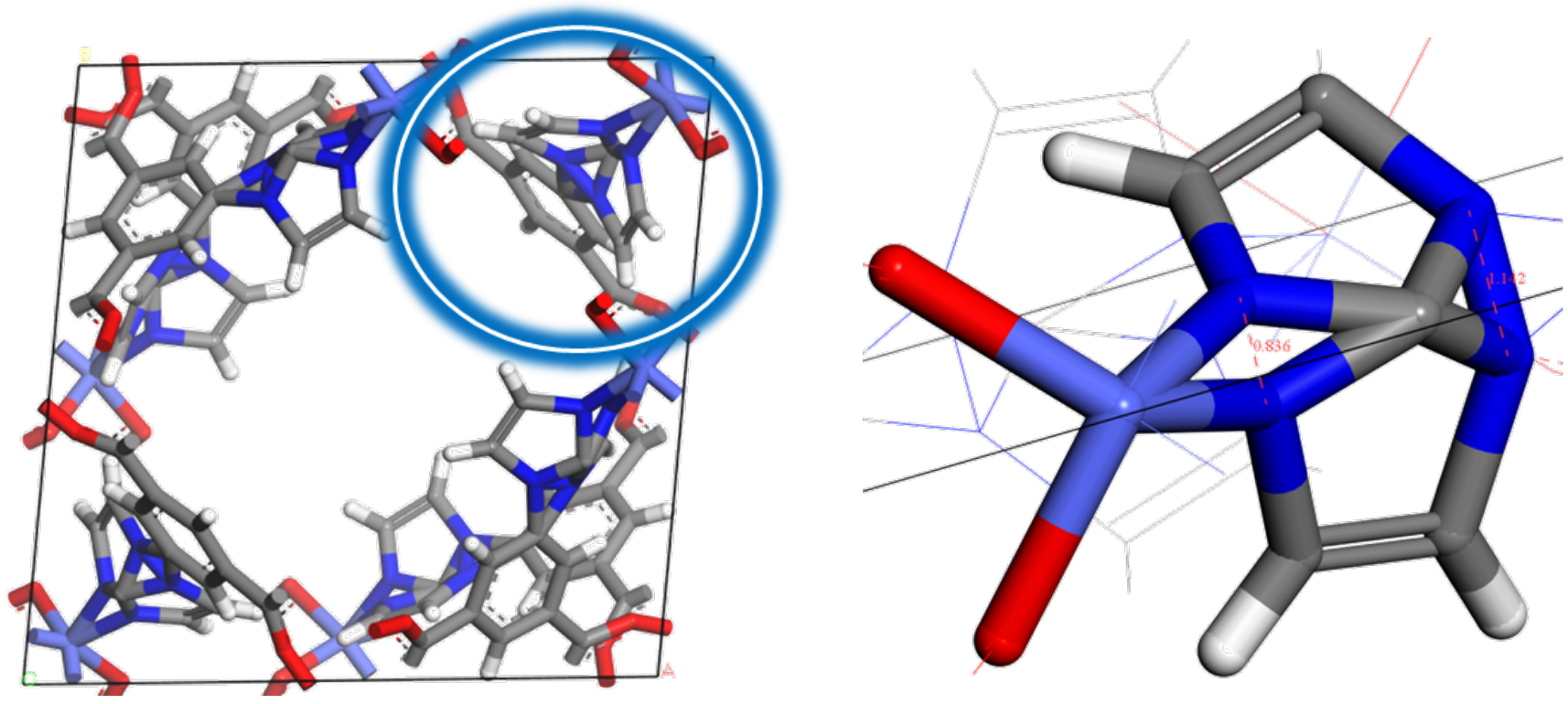

**Figure S5**. Representative NCR structures correctly identified by the “Chen-Manz”.

## S2.2 MOFChecker

For MOFChecker, the most frequently occurring NCR type is “has_overcoordinated_ atom”, with carbon (C) and hydrogen (H) being the most commonly involved elements (**Figure S6**). The type “has_suspicious_terminal_oxo” appears predominantly in FSR structures, likely because CIF files of this type often retain coordinated solvent molecules, which were not removed during curating.

In **Figure S7**, Au-PCM-102 and NTHU-1-Mn were also classified as “free molecule” cases due to bond lengths (e.g., Au-P and Ga-N) exceeding the threshold defined by the algorithm. Since different types of bonds can vary significantly in length-typically, van der Waals bonds > hydrogen bonds > ionic bonds > covalent bonds-it is essential for geometry-based or bond-order-based algorithms to consider the nature of the bond type when determining connectivity. Failure to do so can result in incorrect classifications, such as “under bonded”. In addition, unsaturated alkali metals such as Li or K may exist as free cations in the structure without forming explicit bonds. These cases should not necessarily be interpreted as “undercoordinated_alkali_alkaline”. Similarly, structures flagged as having a “geometrically exposed metal” may, in some cases, represent open metal sites, which are chemically valid. Although MOFChecker2 does not recommend prioritizing the validation of “undercoordinated_ alkali_alkaline” or “geometrically exposed metal” flags, we opted to include them in our screening process to ensure that the positive dataset approaches perfection. Furthermore, for several FSR structures containing “suspicious_terminal_oxo”, we did not remove them because these species were not free solvents. This issue was nearly absent in the ASR structures, which instead exhibited a greater number of “geometrically exposed metal” cases. Details can be found in **Figure S24** of the CoRE MOF 2024 publication.

MOFChecker also performs well in identifying disorder arising from inverted conformers. As shown in **Figure S8**, structures such as FJU-53-Br and M’MOF-20 exhibit rotational disorder of aromatic rings, which is typically flagged as “has_overcoordinated_c”. In contrast, enantiomeric disorder may lead to overcoordinated involving carbon, nitrogen, or hydrogen atoms, and can also be effectively detected by the tool. MOFChecker evaluates whether a structure contains both metal and carbon atoms (as in the case of CoAPO-CJ40, which contains no carbon atoms, further inspection confirmed that it is a molecular sieve rather than a MOF), and whether the pore-limiting diameter (PLD) exceeds 2.4 Å, as part of its criteria to determine whether the structure qualifies as a porous MOF. However, this criterion was not considered in our analysis, since all structures examined in this work have PLDs greater than 2.4 Å. In the case of JUK-14, although the structure exhibits charge imbalance due to the absence of the diethylammonium cation, MOFChecker flagged it under “suspicious_terminal_oxo”, which reflects an issue arising from the lack of solvent removal for ION MOFs in the CoRE MOF DB. However, this is not the primary structural issue in this case.

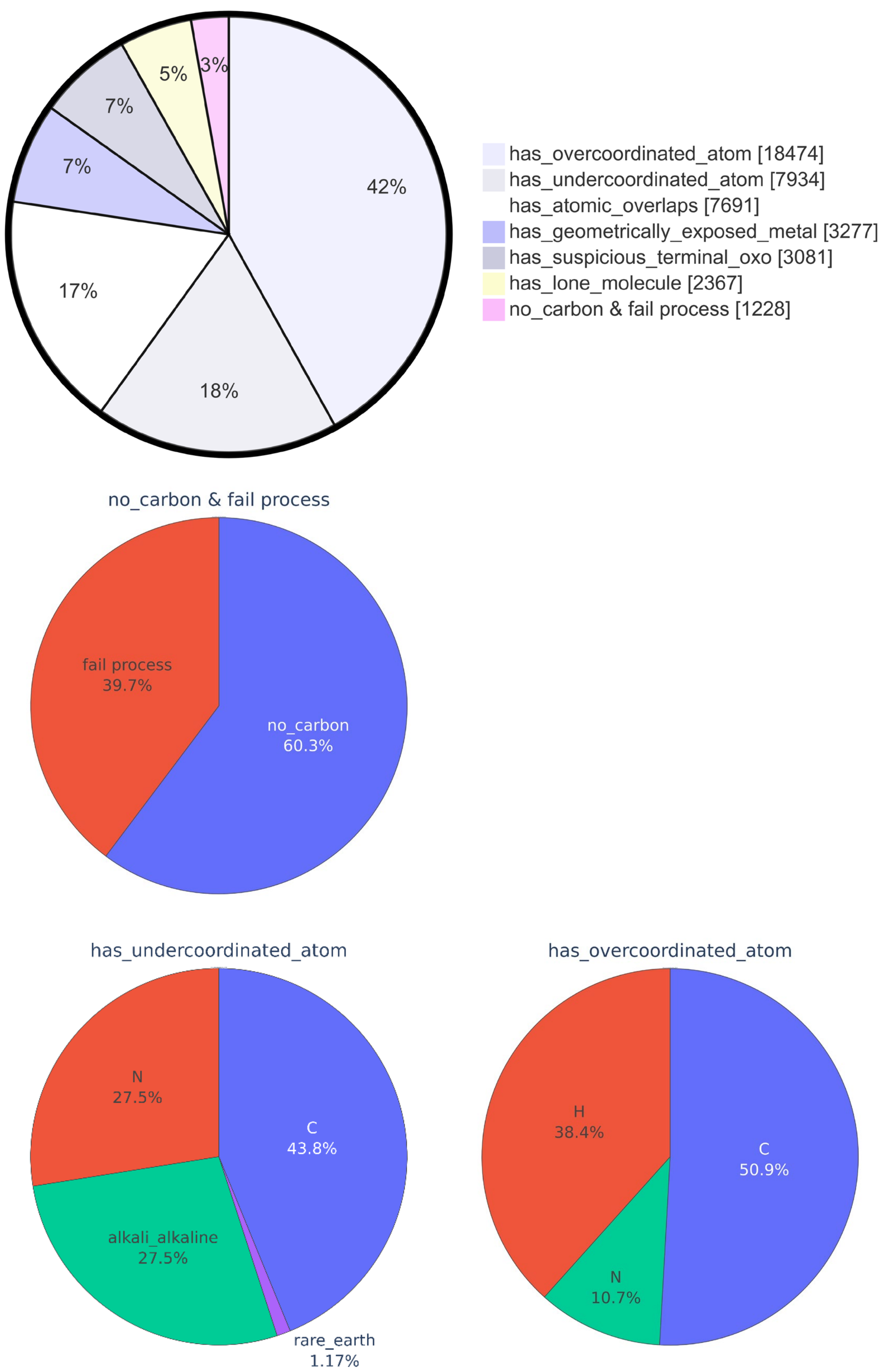

**Figure S6**. Results of each NCR case for CoRE MOF DB defined by MOFChecker. As shown in **Table S1**, we combined the different element with same type of NCR cases (**top**). The distribution of combined cases in the top figure (**middle** and **bottom**).

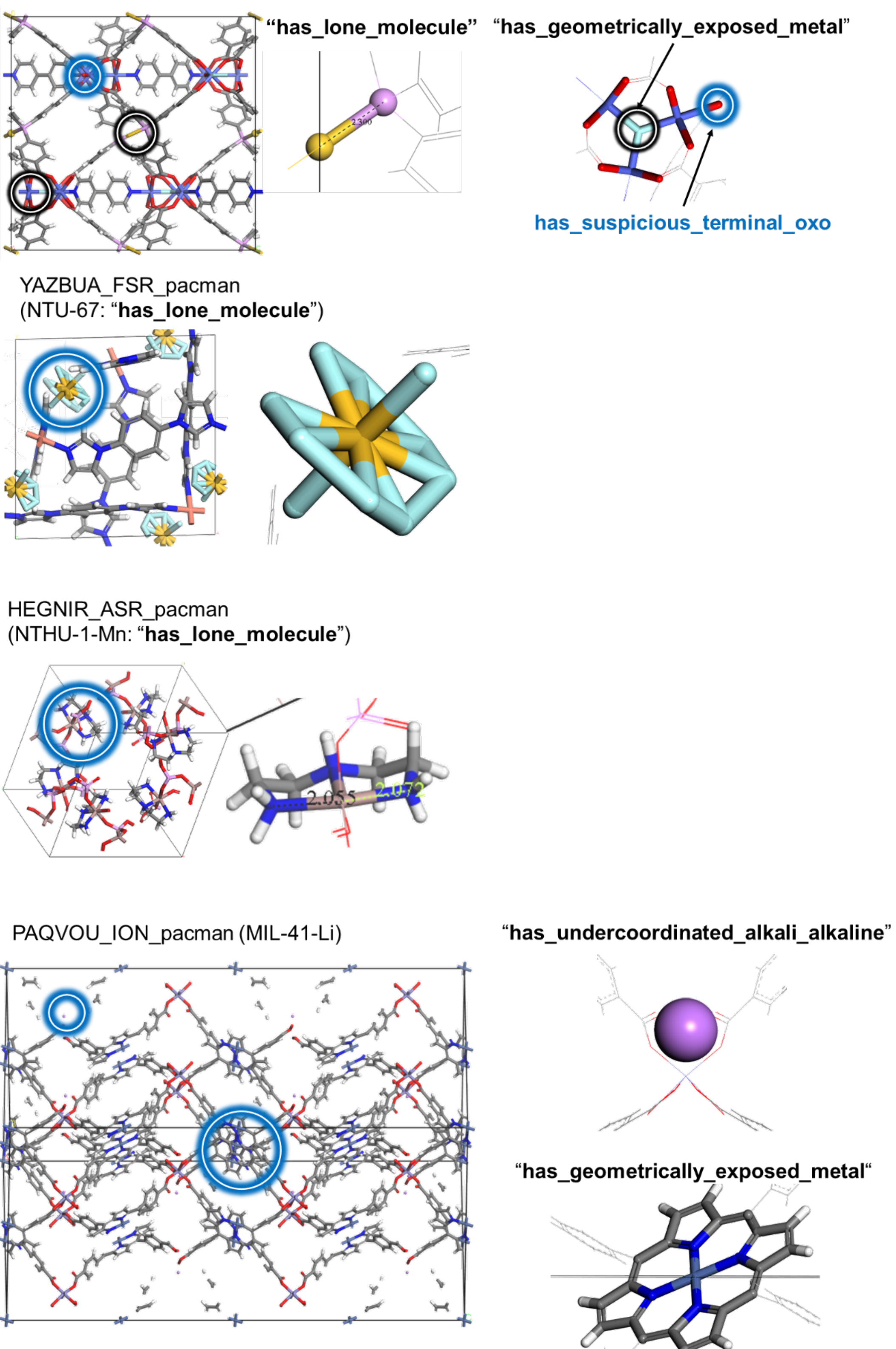

**Figure S7**. Representative NCR structures are misidentified by MOFChecker.

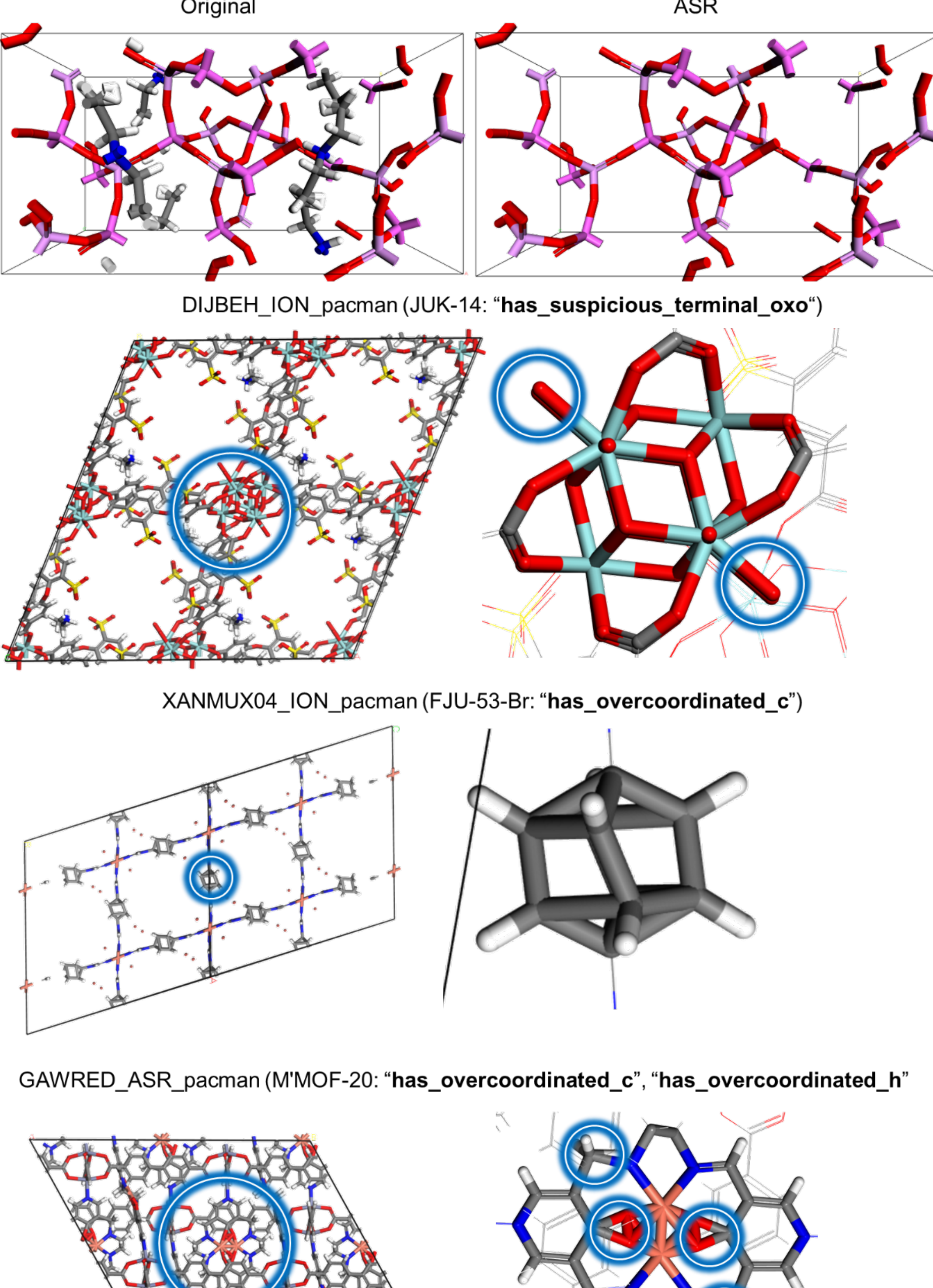

**Figure S8**. Representative NCR structures correctly identified by MOFChecker.

## S2.3 MOSAEC

When categorizing MOSAEC results, the "low_prob_1", "low_prob_2", and "low_prob_3" flags are mutually exclusive by design: a structure assigned "low_prob_3" (probability < 0.01 %) necessarily also meets the criteria for "low_prob_2" (< 0.1 %) and "low_prob_1" (< 1 %). To avoid double-counting, any structure exhibiting multiple "low_prob" flags is recorded only under its most stringent category. Our analysis reveals that the vast majority of structures possess metal oxidation states with occurrence probabilities below 0.01 % in the CSD, implying that these assignments are either fundamentally incorrect or perfect. The high frequencies of "Unknown" and "Impossible" flags further corroborate this interpretation (see **Figure S9**).

Compared to "Chen-Manz" and MOFChecker, MOSAEC shows significantly fewer false positives related to bond-length-based misassignments. However, occasional cases still occur—for example, in **Figure S10**, SUN@PTC-236(Λ) was misclassified as an NCR structure due to one of the $NH_3$ ligands in the $[Zn(NH_3)_4]^{2+}$ unit being positioned slightly far from the Zn center. In addition, MOSAEC occasionally misclassifies structurally CR structures as NCR, such as MOF-217 and DMOF-13. MOSAEC performs poorly in cases such as MOF-20 (**Figure S8**) and ECUT-111 (**Figure S5**), where the metal oxidation states are correctly represented despite the presence of rotational disorder of aromatic rings or inverted conformers in the linkers. In such situations, MOSAEC fails to accurately identify the structural disorder.

However, MOSAEC excels at detecting missing ions required for charge balance in the framework, as demonstrated in **Figure S11** by structures like PCN-9 (Mn) (missing $H_3O^+$) and $[Cd_4(bpt)_2(suc)_{2.5}(OH)]\cdot 3H_2O$ (missing $OH^-$). This strength stems from MOSAEC's underlying algorithm, which evaluates each linker atom coordinated to a metal center and calculates the oxidation state of the metal. The metal oxidation state provides indirect insight into the net framework charge, allowing accurate identification of charge imbalance. Furthermore, MOSAEC is also effective at identifying incorrect protonation states. For example, in BAKGIF,[6] the presence of an incorrect number of protons, and in UiO-67, the absence of hydrogen atoms on $\mu_3$-oxo bridges between metal centers, were both successfully flagged. Once BAKGIF was corrected, MOSAEC was also able to classify the revised structure as "GOOD".

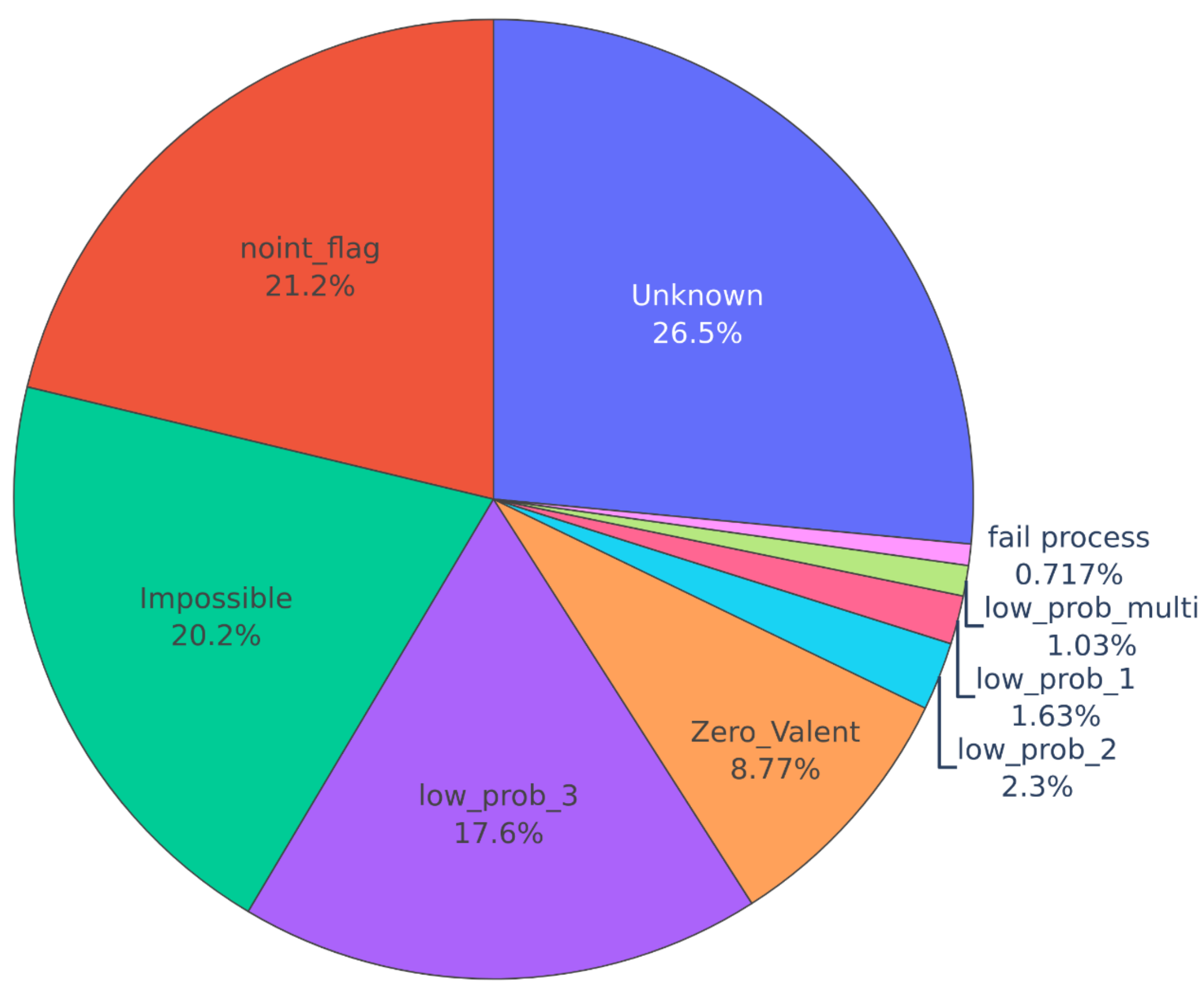

**Figure S9**. Results of each NCR case for CoRE MOF DB defined by MOSAEC.

UDOJOQ_ION_pacman (SUN@PTC-236(Λ): "**Impossible**", "**Unknown**", "**Zero_Valent**", "**noint_flag**", "**low_prob_1**", "**low_prob_2**", "**low_prob_3**", "**low_prob_multi**")

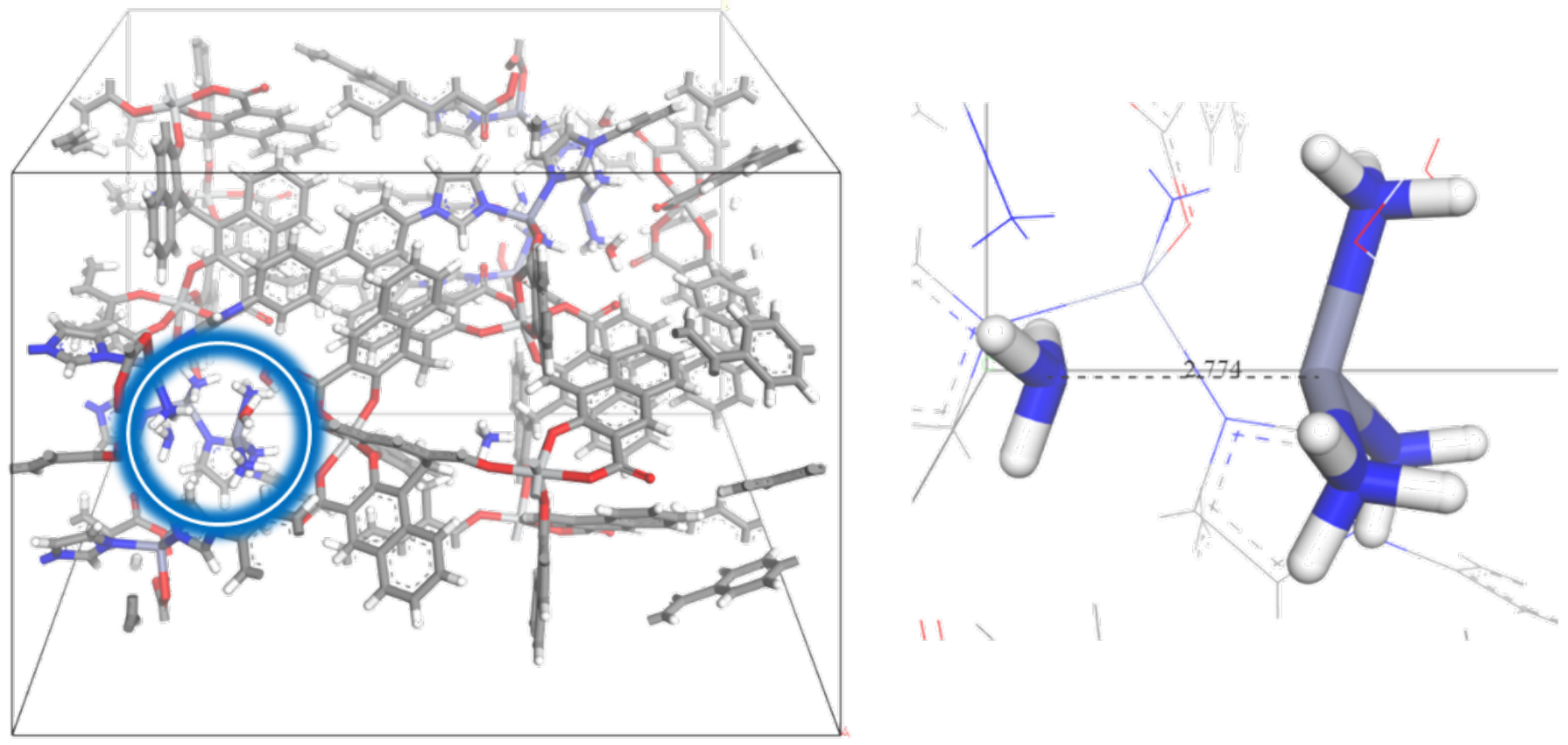

c9sc06500h2_ASR_pacman (MOF-217: "**Impossible**", "**Unknown**", "**low_prob_1**", "**low_prob_2**", "**low_prob_3**")

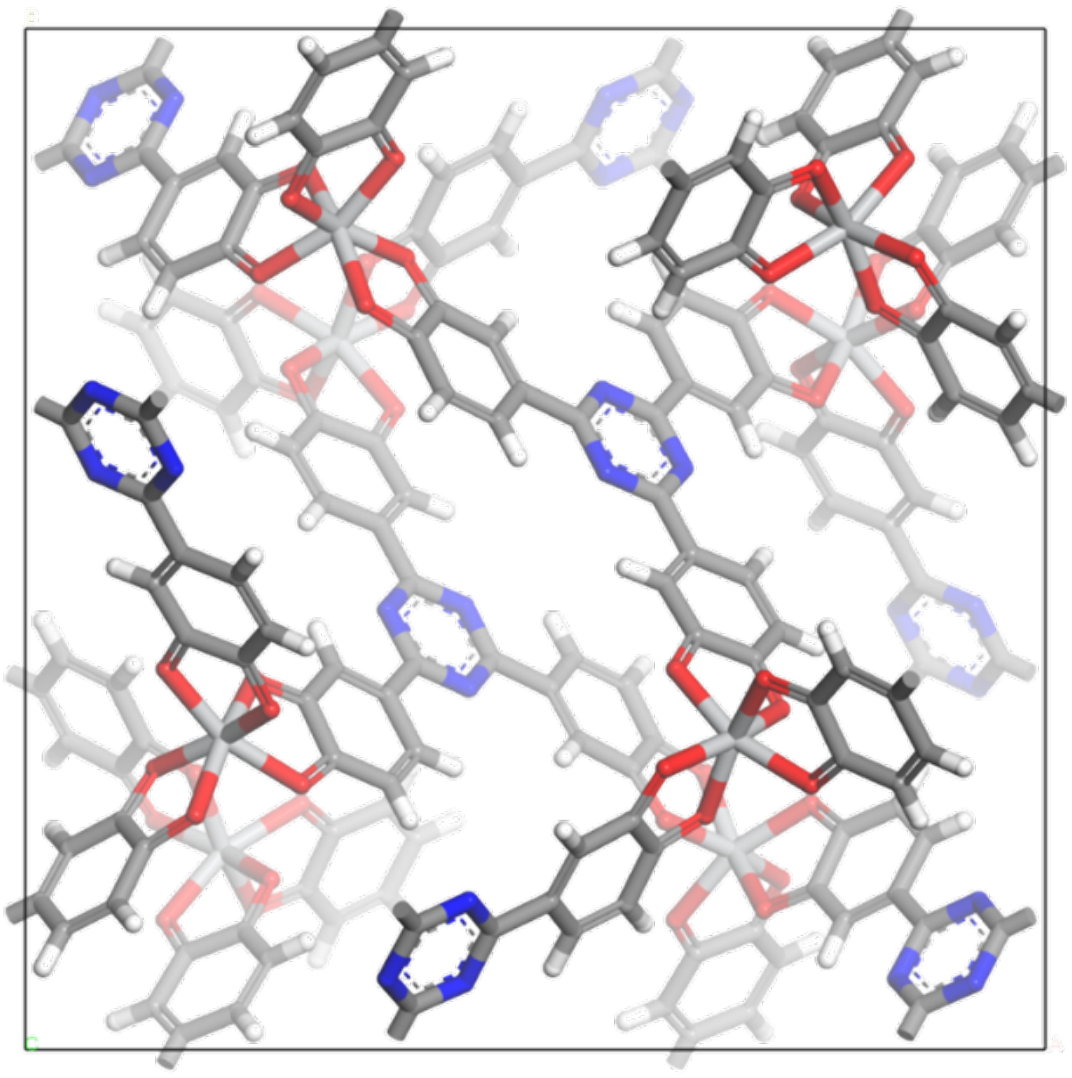

FAYRAA_FSR_pacman (DMOF-13: " **noint_flag**")

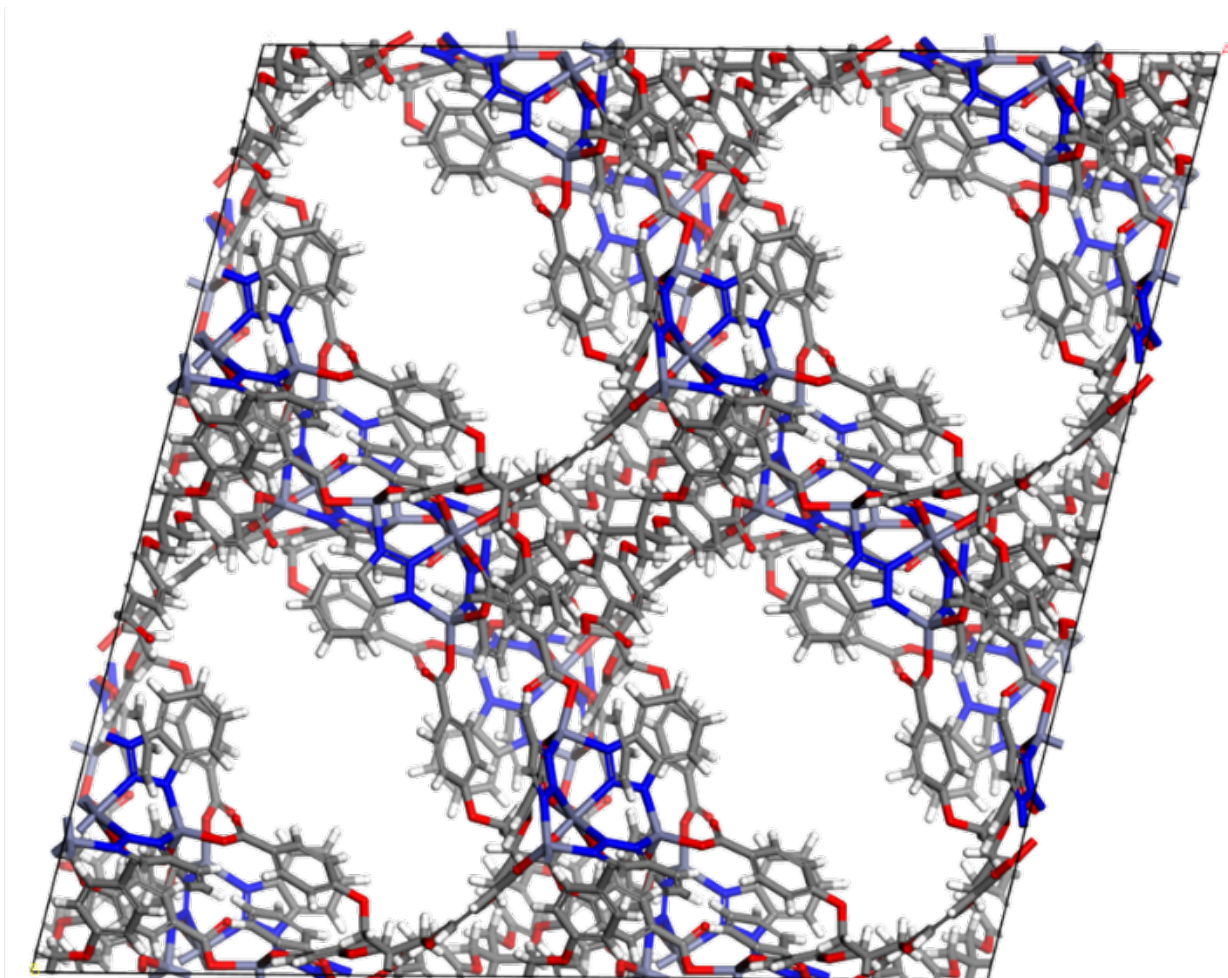

**Figure S10**. Representative NCR structures are misidentified by MOSAEC.

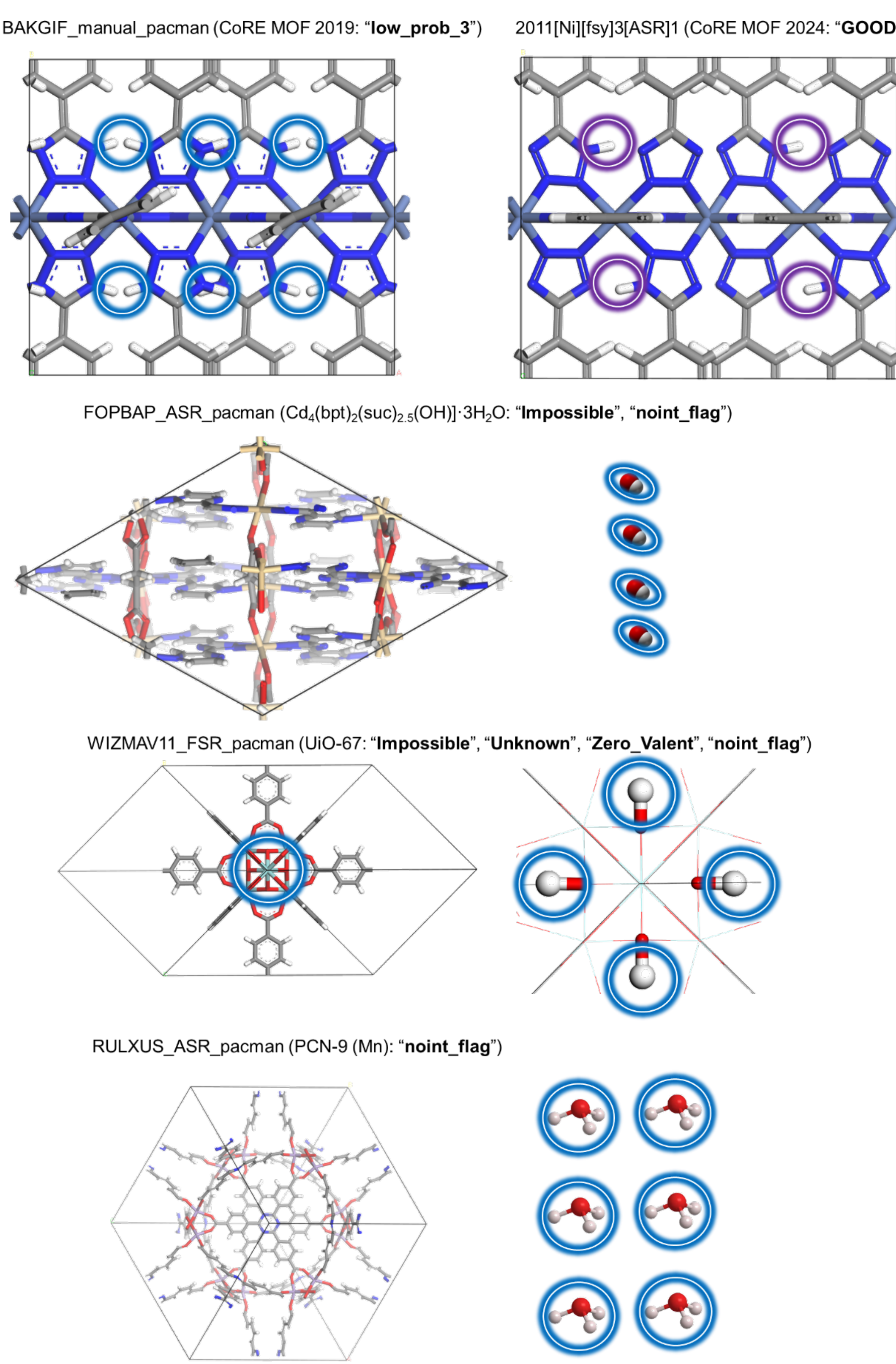

**Figure S11**. Representative NCR structures correctly identified by MOSAEC.

## S3 Graph Neutral Network

We compared the atomic charge of CR MOF and NCR MOF in our previous study, and we also found that the PACMAN model[7] (GCN-based) can predict partial atomic charges of ions in the ION MOFs accurately. As shown in **Figure S12**, we incorporated PACMAN into the solvent removal algorithm to detect whether free or coordinated molecules are ions. This approach eliminates the need to define a fixed ion list to identify removable species, thereby avoiding the unintended removal of ions that are not explicitly included in the list. The results, presented in **Figure S13**, demonstrate that the GCN-based model can successfully predict the charges of both free and coordinated ions, without interference from solvents such as $H_2O$ or DMF (BOMCUB, YARGIJ, BUJLOK). The model is also effective in cases involving multiple types of ions (ZAFRUX) or ions with a larger number of atoms (XUVGUQ). Furthermore, the presence of disorder in ligands or metal nodes does not compromise ion identification (AFEJOK), and even $H_2O$ molecules lacking hydrogen atoms (only oxygen present) are predicted with charges close to zero. However, it is important to note that when the ions themselves have disorder, charge prediction becomes unreliable. For example, in APUQEK, a disorder anion ($SbF_6 \rightarrow Sb_2F_{12}$) was incorrectly predicted to carry a positive charge. This indicates that GCNs may yield physically unrealistic predictions when structural disorder is present. Therefore, using GCN models to assess overall charge neutrality or structural disorder in MOFs remains a feasible approach. The code is available in Github (https://github.com/mtap-research/CoRE-MOF-Tools/blob/main/CoREMOF/curate.py).

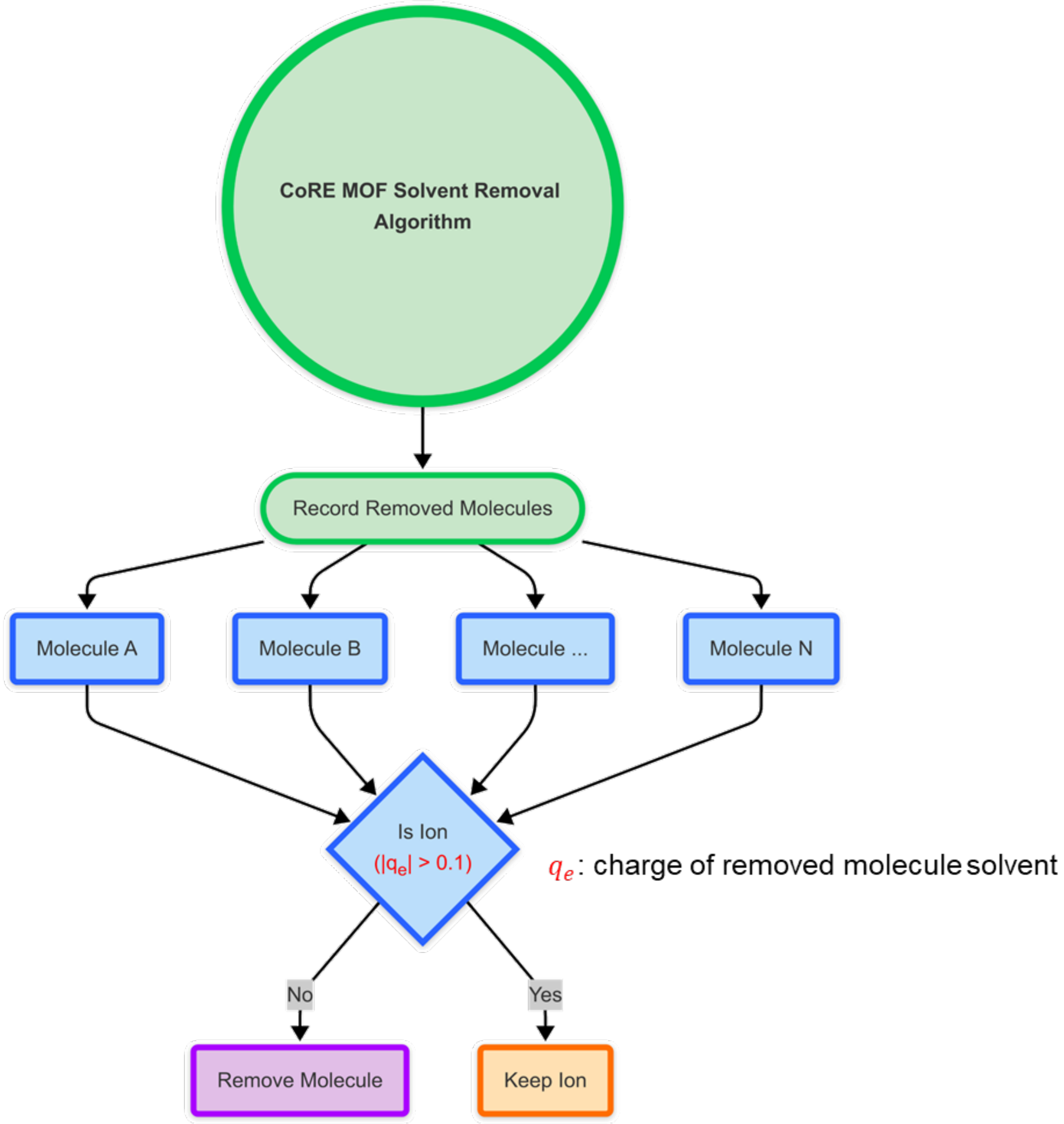

**Figure S12**. The workflow of ion charge checking by PACMAN and curation algorithm based on CoRE MOF work.

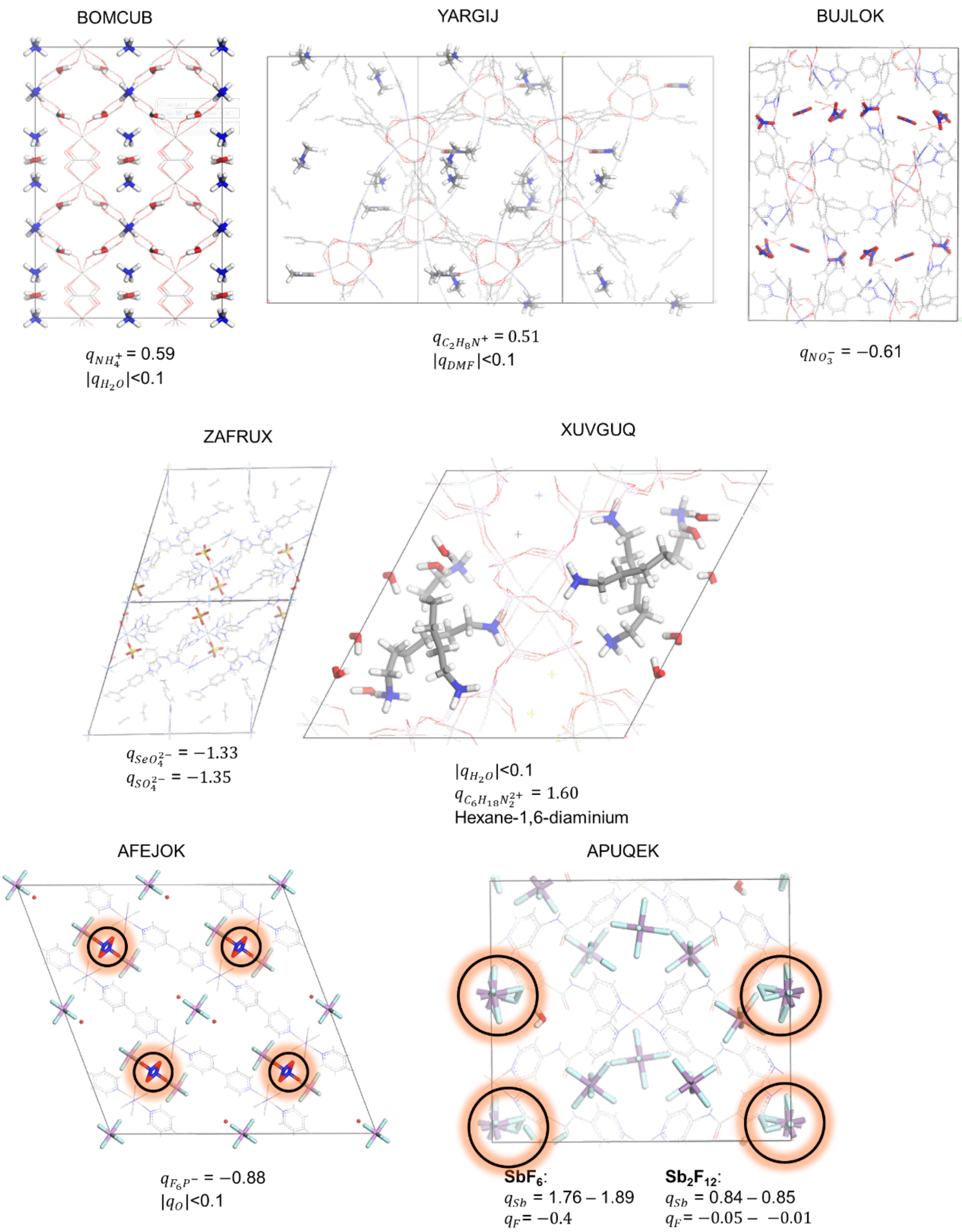

**Figure S13**. The predicted charges of ions by GCN-based model (PACMAN).

# S4 PU-CGCNN Model Training

## S4.1 Crystal Graph Convolutional Neural Network

We converted each structure into a crystal graph representation (with nodes and edges) using ASE and pymatgen, with a cutoff radius of 8 Å for neighbor searching and a maximum of 12 neighbors per atom. We employed the CrystalNN algorithm by Pymatgen to determine the coordination numbers of metal centers across MOFs from CoRE MOF 2025. As shown in **Figure S14**, we found that the coordination numbers of all metal centers-except for Ta and Cs-are generally within 12 and the longest observed bond length is 3.94 Å; thus, a cutoff of 8 Å is sufficient to capture the indices of both first and second coordination atoms while avoiding excessive inclusion of irrelevant long-range interactions. The same settings are also used in other related studies of CGCNN models applied to MOFs.[8-12] For more details on the CGCNN architecture, please refer to the work by Xie et al. and related studies such as PACMAN. Structures that could not be successfully converted into graph representations were classified as NCR and excluded from the training process. Furthermore, any structure that failed during graph construction was defined as NCR by default.

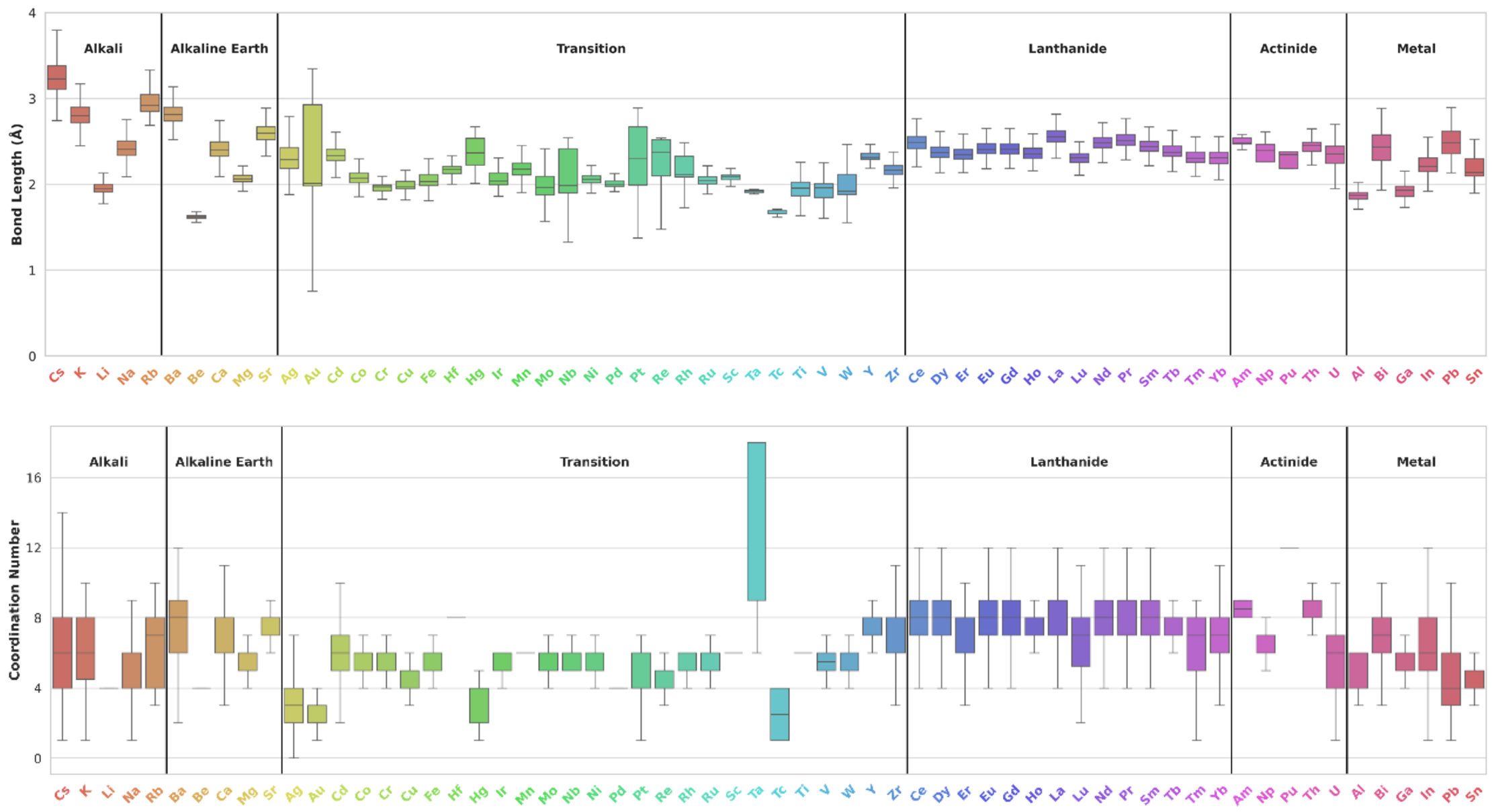

**Figure S14**. The bond length (Top) and coordination number (Bottom) of different types of metal for CoRE MOF 2025. We employed the CrystalNN algorithm from the pymatgen library to determine the coordination numbers of metal centers across all structures.

## S4.2 PU Learning

We trained our model using 100 iterations of Bootstrap Aggregating (bagging),[13] an ensemble learning technique well-suited for models that are sensitive to data perturbations. In each iteration, a subset of the data was randomly sampled with replacement to train an individual model, and the final prediction was obtained by averaging the outputs from all models. Each iteration was trained for 30 epochs, during which we randomly (in each iteration, a new selection was made to ensure that the 100 bagging models collectively covered all structures in the dataset with diverse combinations) selected 13,213 unlabeled samples—equal to the number of positive samples—and temporarily labeled them as negative data. To evaluate model performance, 20% of both the positive and unlabeled sets were held out as a validation set. The final structure score for each sample in the test set was calculated as the average prediction score from the CGCNN classifiers across all iterations. The predicted scores range from 0 to 1, where a score of 1 indicates a structurally perfect MOF, and a score closer to 0 reflects a higher degree of disorder. The goal of this model is to estimate how good a structure is in terms of its crystallographic quality.

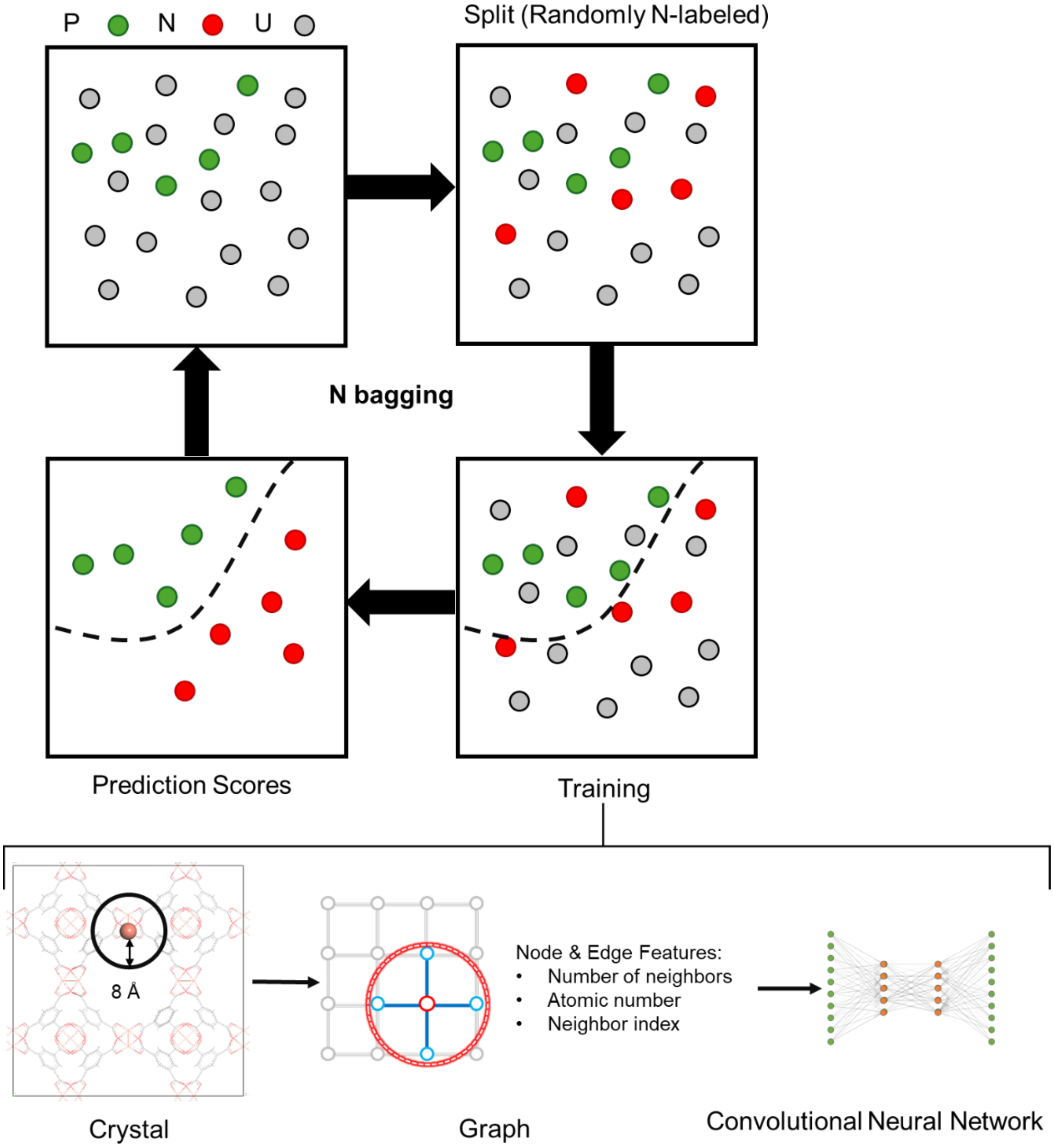

**Figure S15**. Data partitioning, training workflow, and iterative procedure (bagging) based on the PU-CGCNN (P: positive; N: negative; U: unlabeled). Schematic diagram of the CGCNN architecture.

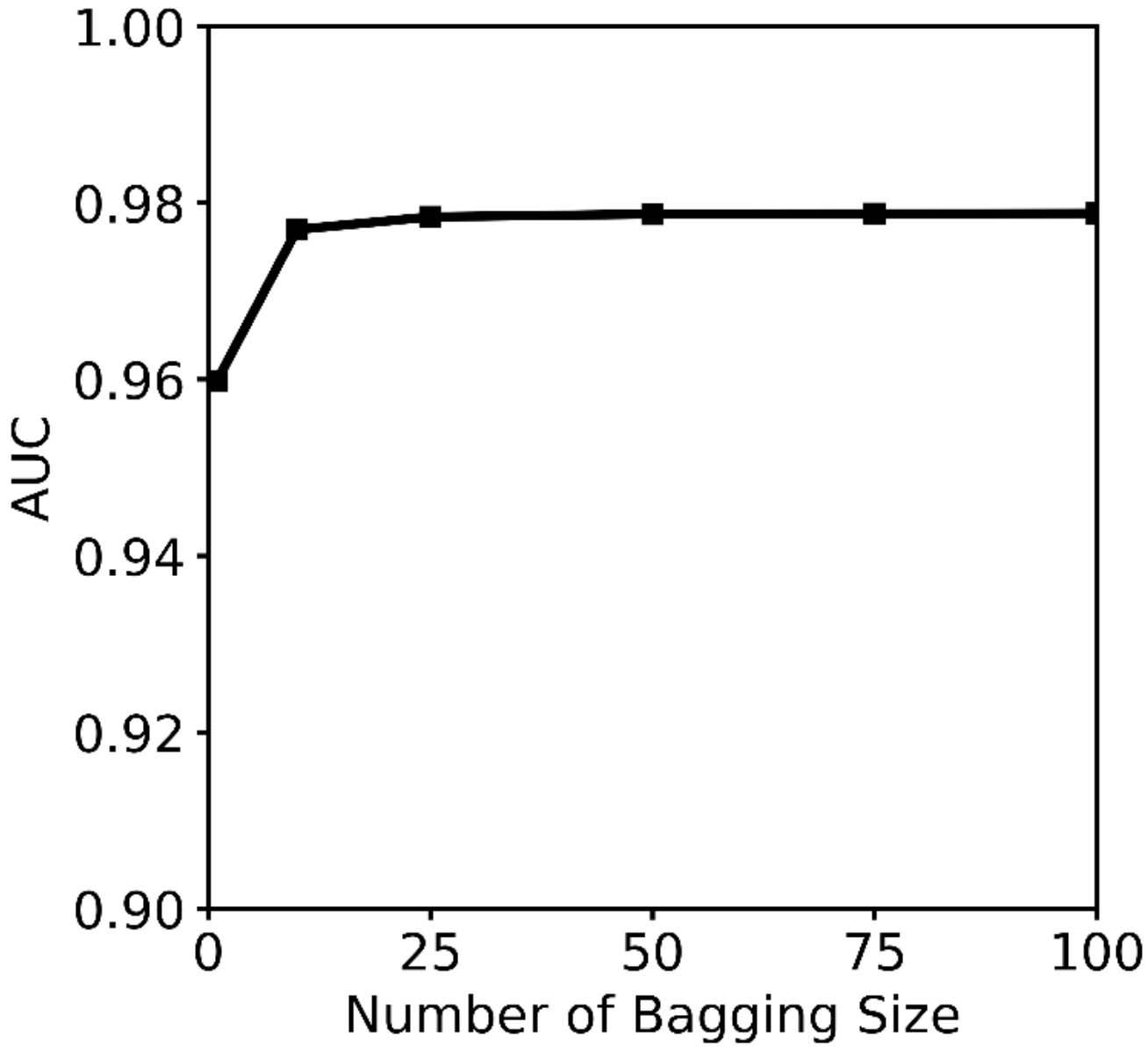

**Figure S16**. The convergence of the model's AUC (area under the curve) was enhanced by increasing the bagging size from 1 to 100.

**Table S2**. The recall rate, precision, and F1 score of MOFClassifier and number of CR structures (43,439 MOFs in total) with different cutoff values. The recall rate, precision, and F1 score of other checkers for CR and NCR classification The definition of "All" here differs from that in **Figure 2**, in the table, structure is labeled as positive only if it passes all three checkers simultaneously; otherwise, it is labeled as negative. The equation can be found in **Eq. S1**, **Eq. S2**, and **Eq. S3**.

| Checker | Cutoff | Recall | Precision | F1 | CR |
|---|---|---|---|---|---|
| MOFClassifier | 0.1 | 0.999 | 0.579 | 0.733 | 23,380 |
| | 0.2 | 0.996 | 0.647 | 0.784 | 20,805 |
| | 0.3 | 0.991 | 0.703 | 0.822 | 19,014 |
| | 0.4 | 0.980 | 0.759 | 0.856 | 17,407 |
| | 0.5 | 0.961 | 0.806 | 0.877 | 16,036 |
| | 0.6 | 0.933 | 0.850 | 0.890 | 14,725 |
| | 0.7 | 0.885 | 0.896 | 0.890 | 13,221 |
| | 0.8 | 0.781 | 0.939 | 0.853 | 11,135 |
| | 0.9 | 0.508 | 0.985 | 0.670 | 6,909 |
| All | – | 0.989 | 0.556 | 0.712 | 13,384 |
| Chen-Manz | – | 0.946 | 0.624 | 0.752 | 24,995 |
| MOFChecker | – | 0.923 | 0.709 | 0.802 | 21,463 |
| MOSAEC | – | 0.888 | 0.727 | 0.799 | 20,010 |

$$Recall = \frac{TP}{TP + FN} \quad (S1)$$

$$Precision = \frac{TP}{(TP + FP)} \quad (S2)$$

$$F1 = 2 \times \frac{Precision \times Recall}{Precision + Recall} \quad (S3)$$

where TP, FP, and FN are true positive, false positive, and false negative.

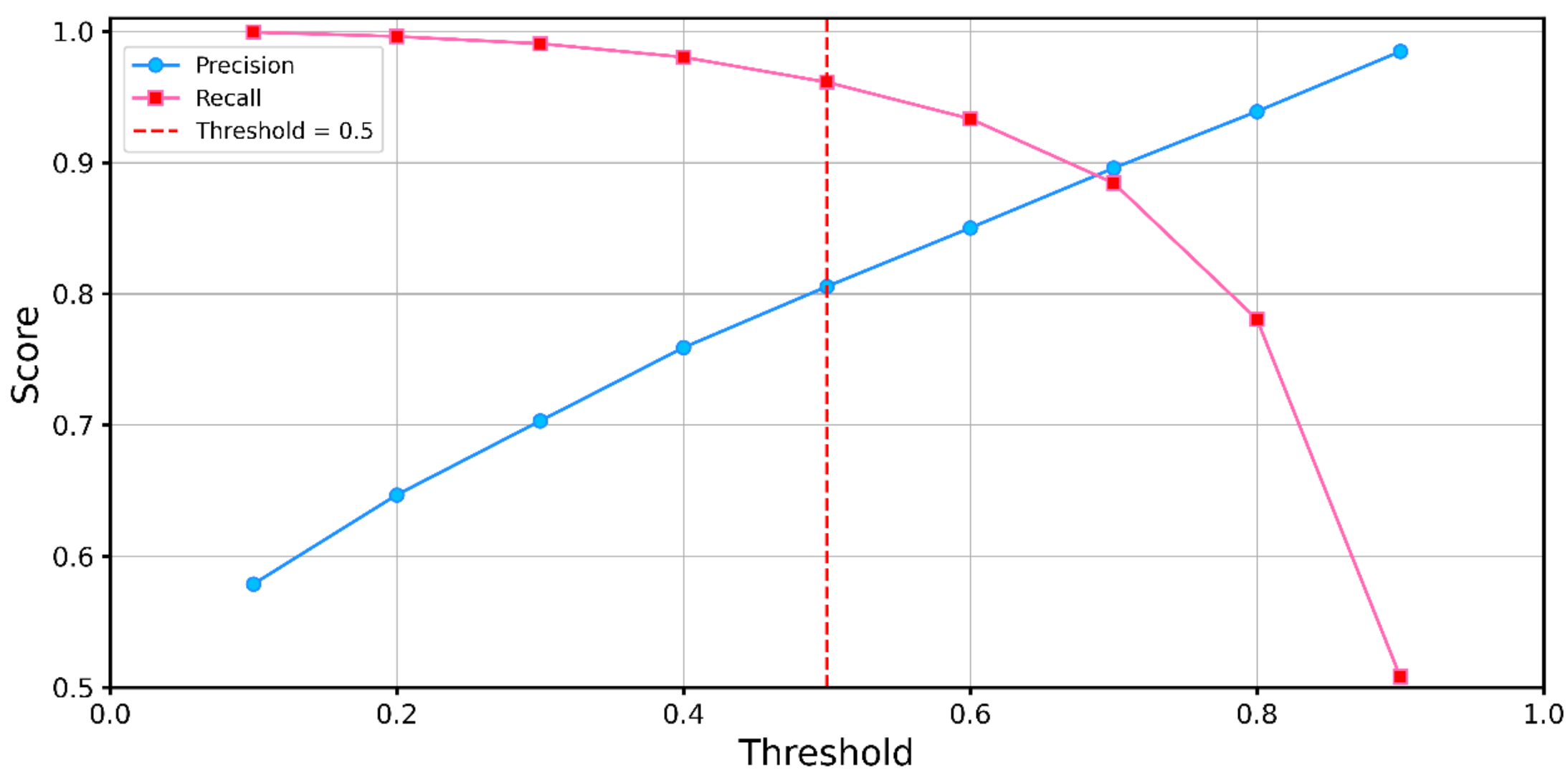

**Figure S17**. Precision and recall as a function of the classification threshold based on 5-fold cross-validation. The red dashed line indicates the default threshold of 0.5

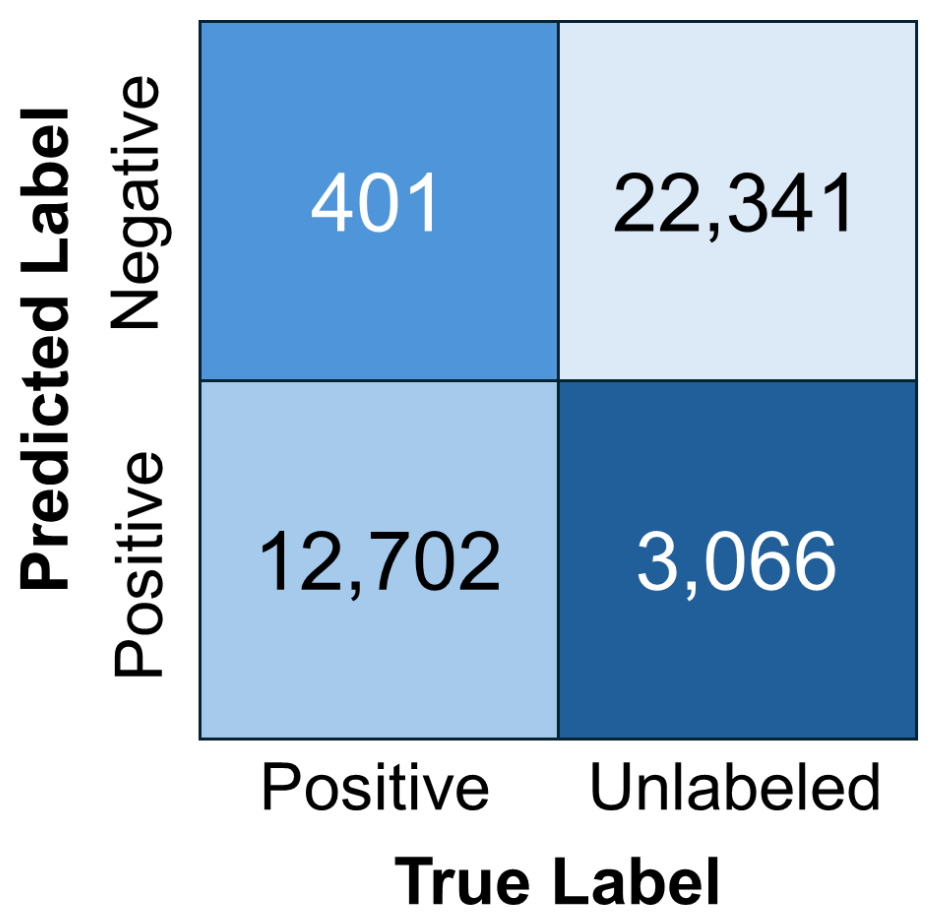

**Figure S18**. Confusion matrix of CLscore predictions of MOFClassifier. The positive structures defined by MOFClassifier, "Chen-Manz", MOFChecker and MOSAEC.

## S5 Result of MOFClassifier

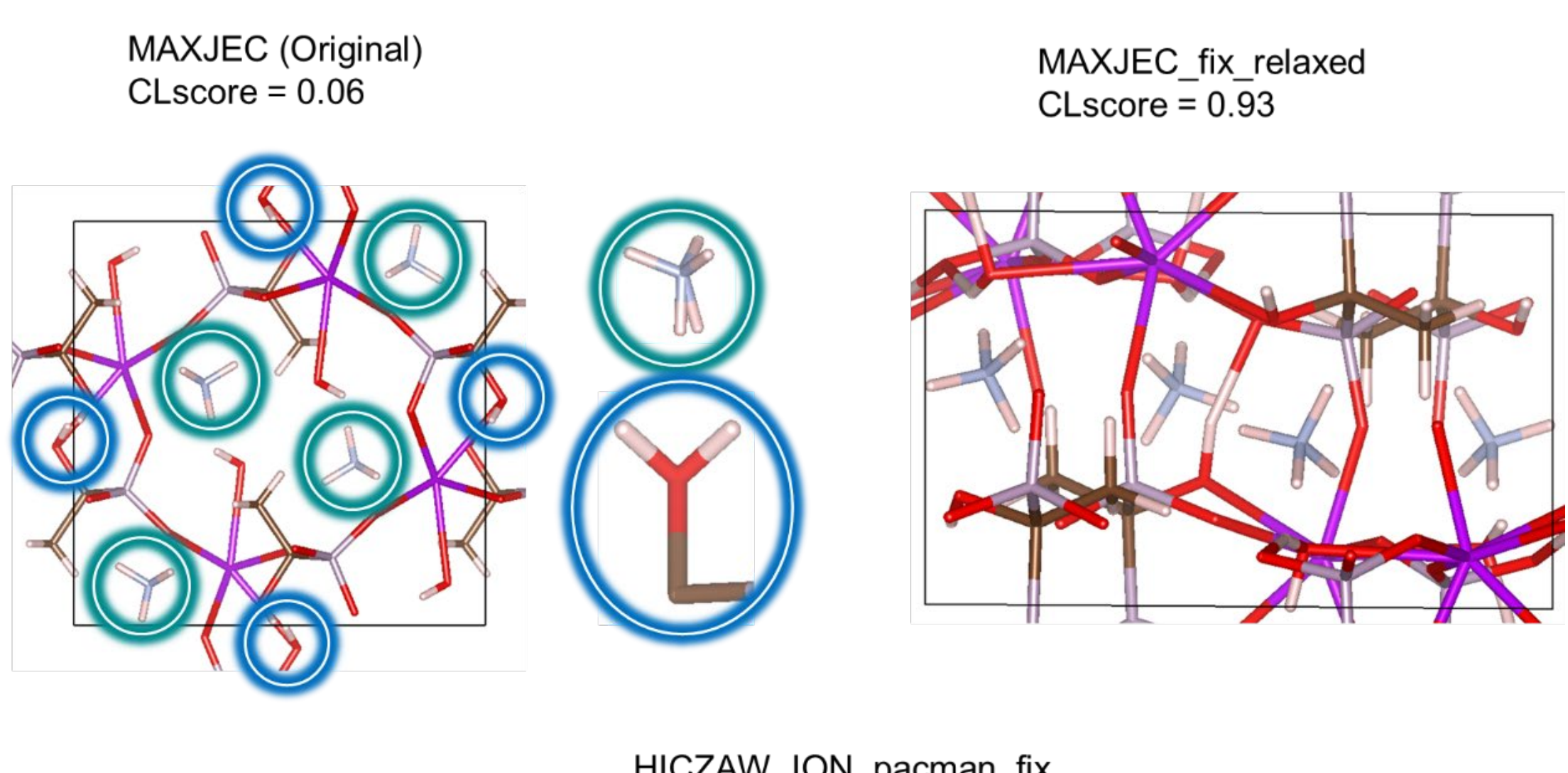

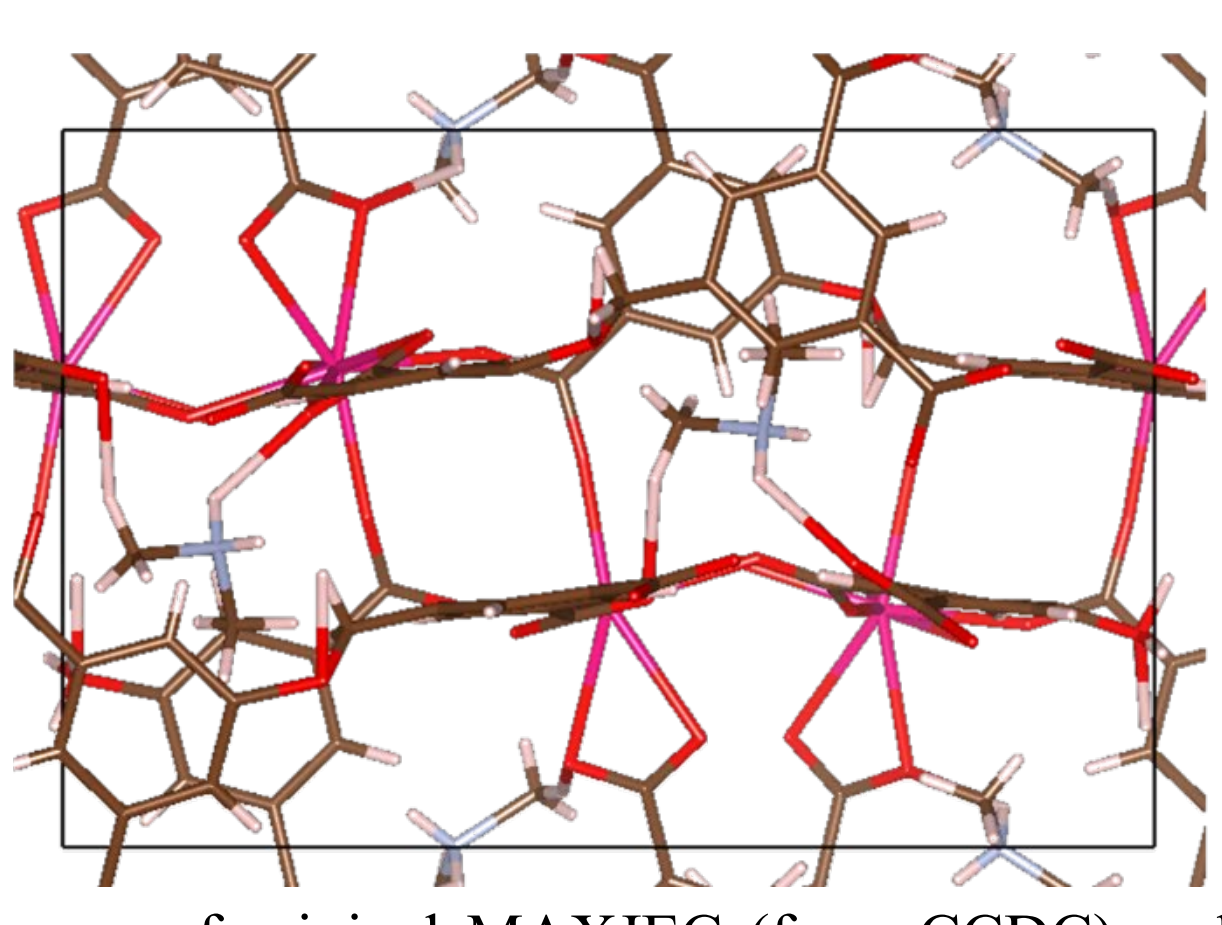

**Figure S19**. The CLscore of original MAXJEC (from CCDC) and fixed MAXJEC (MAXJEC_fix_relaxed) predicted by MOFClassifier (**Top**). We fixed the MAXJEC based on reported paper and relaxed the structure by VASP 6.4.3 (Cell relaxation was performed using the PBE exchange-correlation functional with D3-BJ dispersion correction, a plane-wave cutoff energy of 520 eV, and spin polarization. The SCF electronic convergence threshold was set to 10E-6 eV, and the ionic relaxation convergence criterion was 0.02 eV/Å for the maximum force. The maximum number of ionic steps was set to 200. A Γ-centered 3×2×3 k-point mesh was used for Brillouin zone sampling). The CLscore of HICZAW_ION_pacman_fix predicted by MOFClassifier (**Bottom**).

### S5.1 False Negative by Three Methods

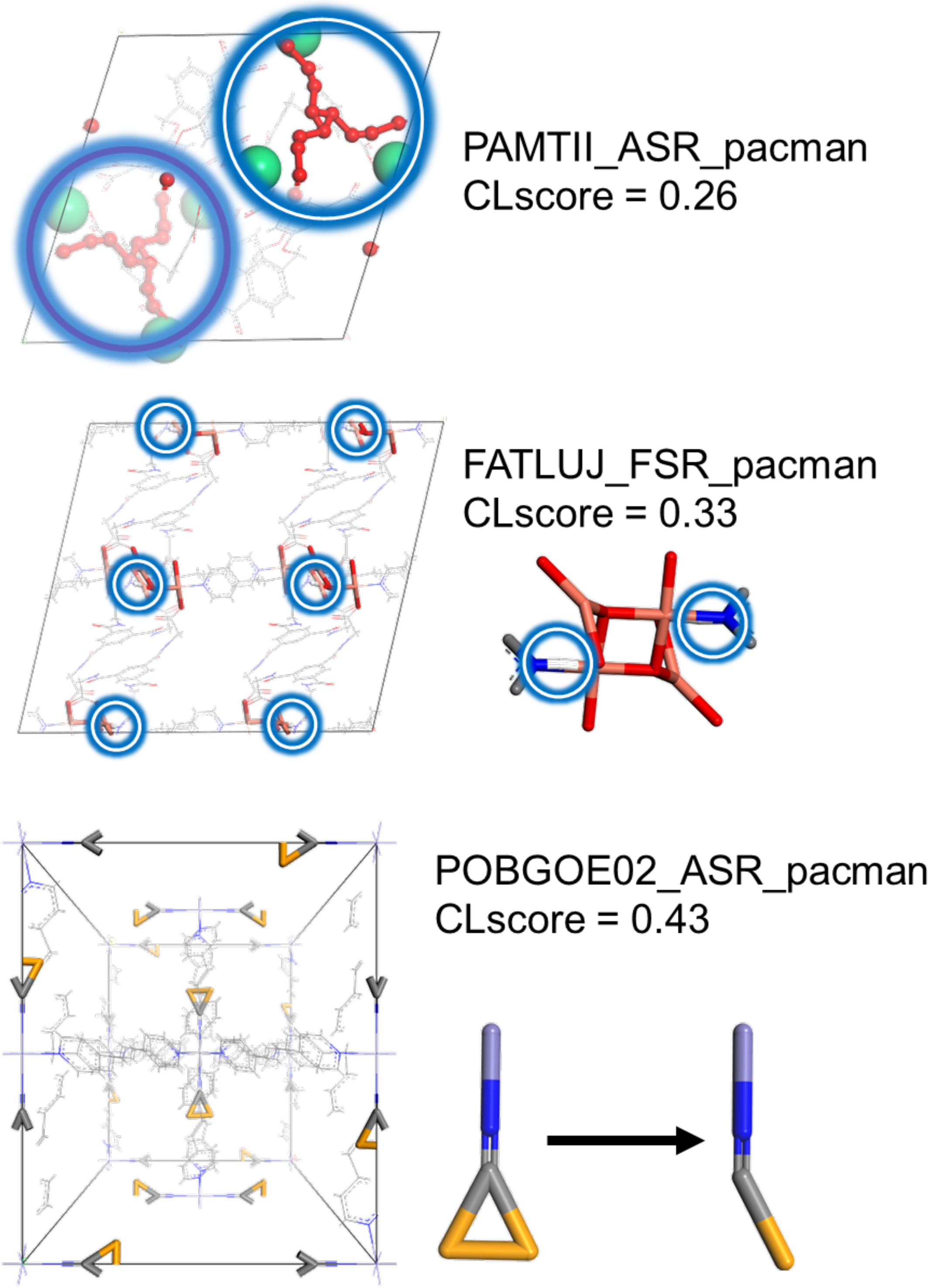

**Figure S20**. Representative NCR structure cases with apparent disorders from the positive sample set that were predicted as NCR structures with CLscores below 0.5.

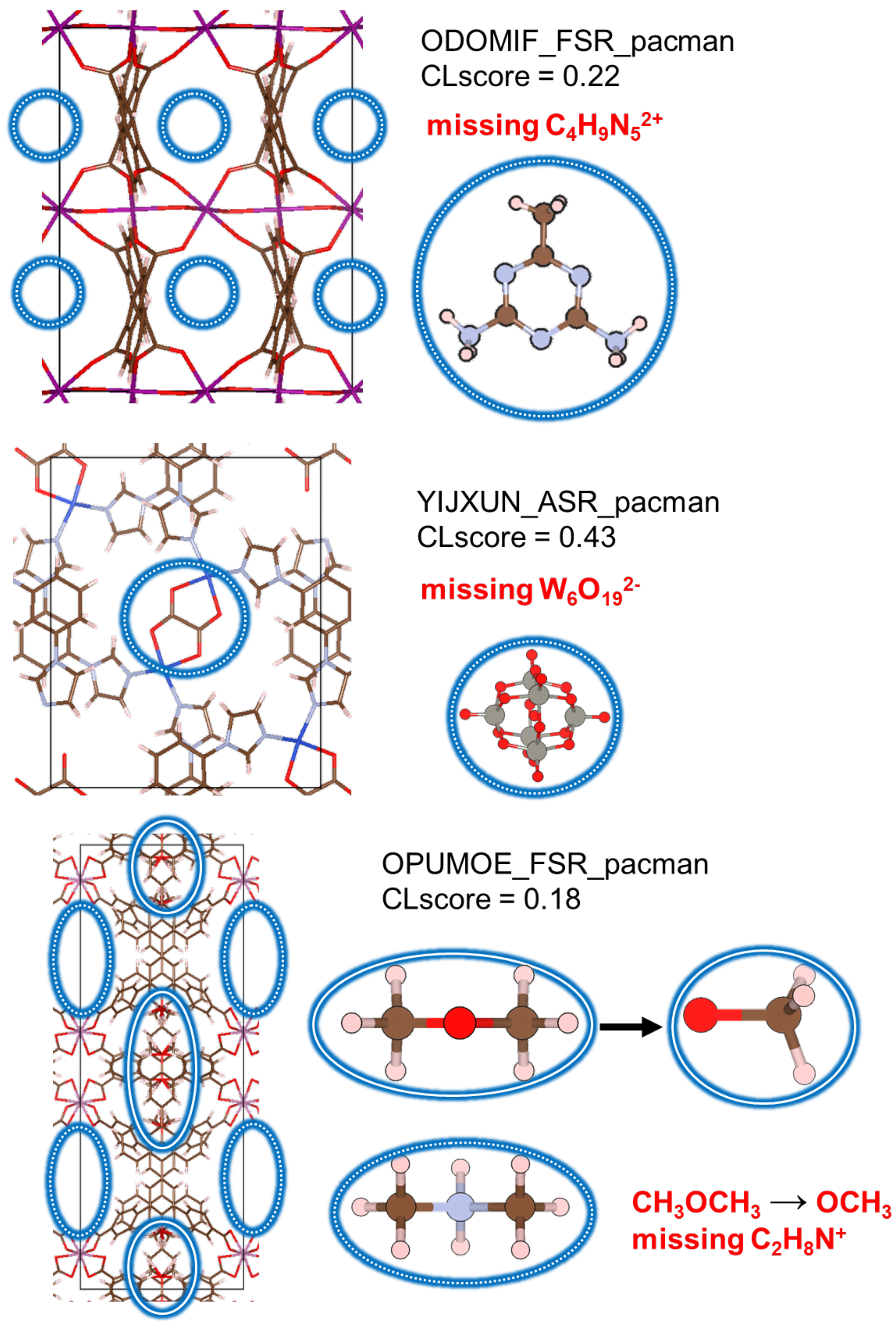

**Figure S21**. Representative NCR structure cases with missing ions from the positive sample set that were predicted as NCR structures with CLscores below 0.5.

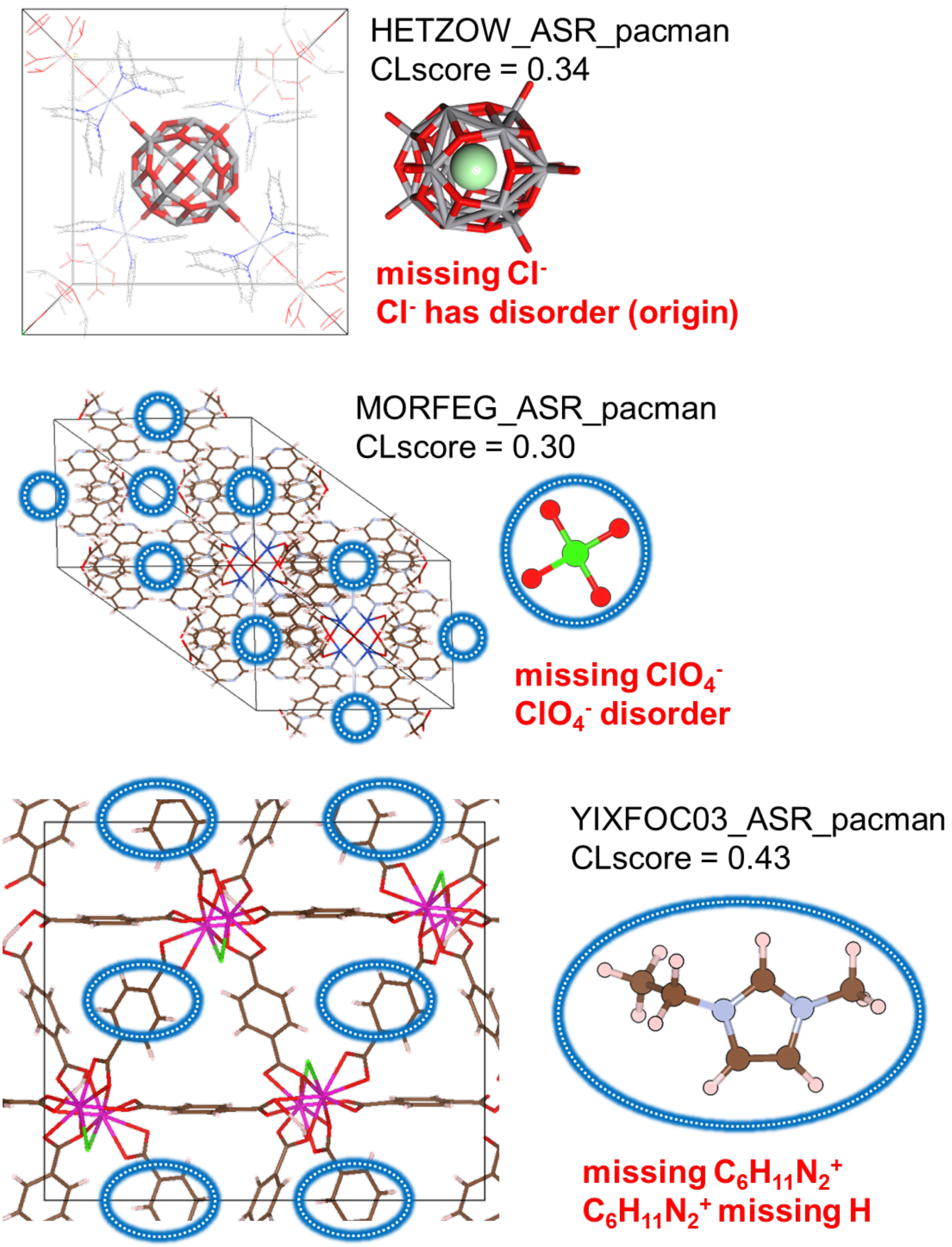

**Figure S22**. Representative NCR structure cases with missing ions (ions removed due to ions have disorder) from the positive sample set that were predicted as NCR structures with CLscores below 0.5.

## S5.2 False Positive by Three Methods

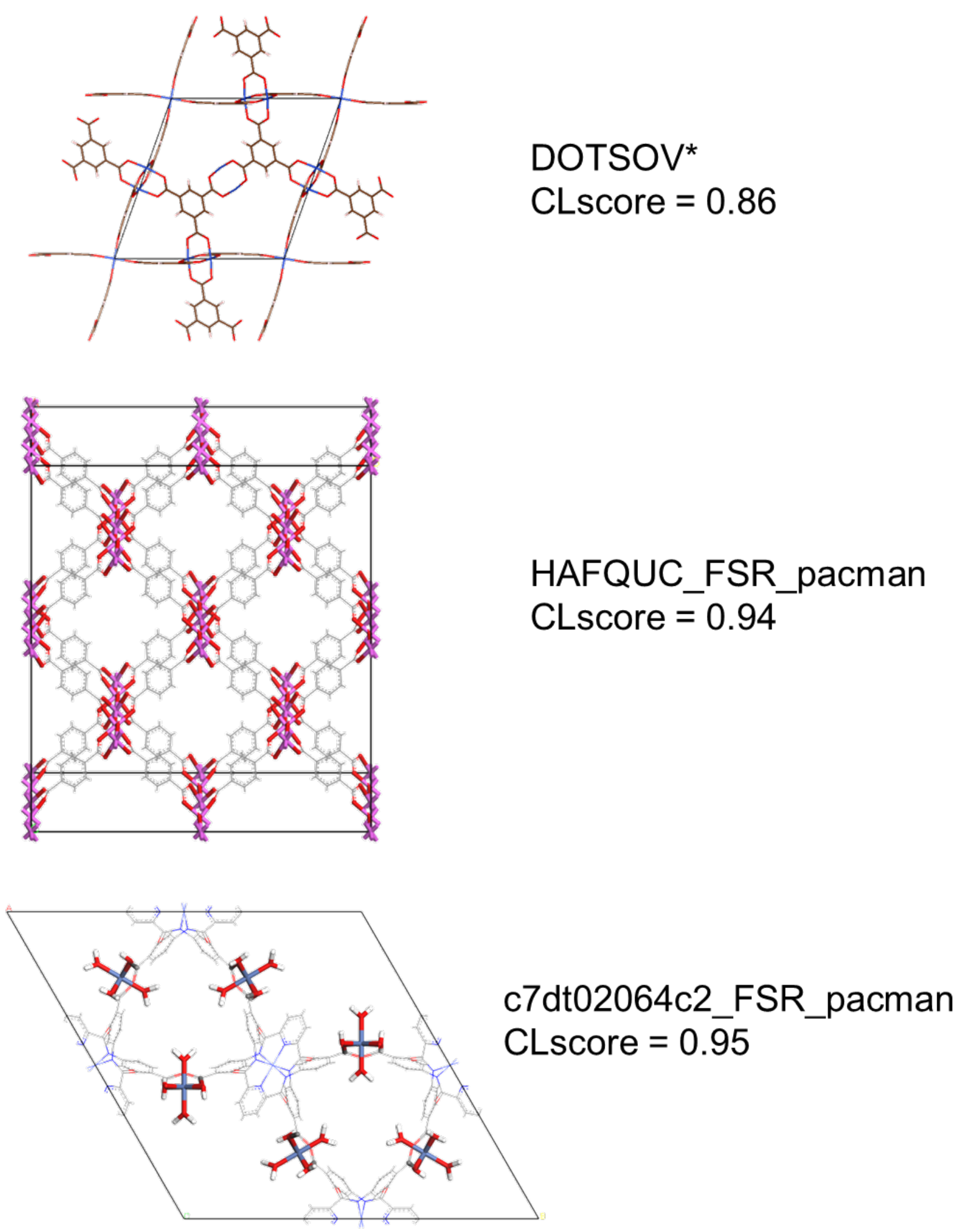

**Figure S23**. Representative CR structure cases from the unlabeled sample set that were predicted as CR structures with CLscores above 0.5.

## S5.3 False Negative by MOFClassifier

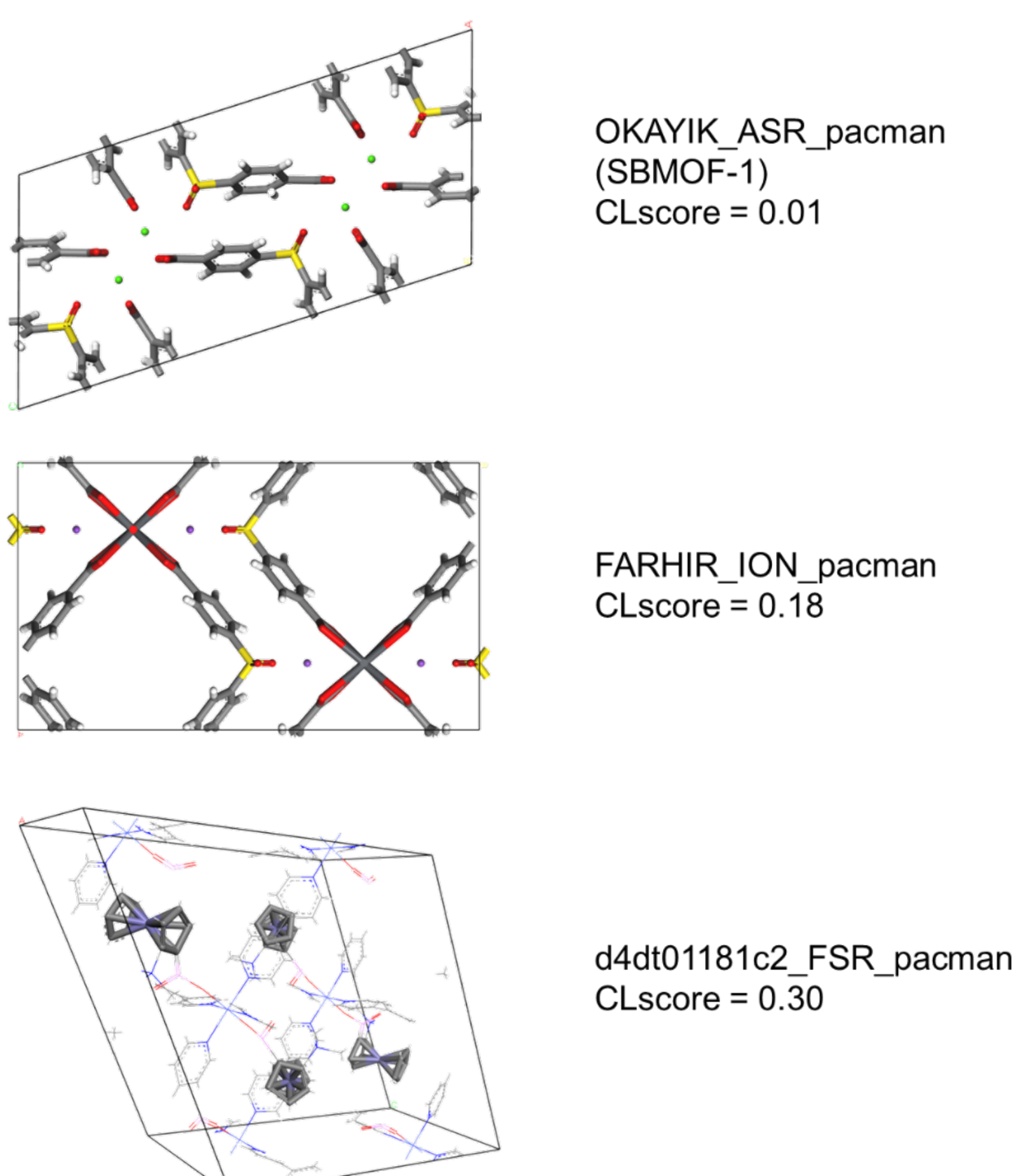

**Figure S24**. Representative CR structure cases from the positive and unlabeled sample set that were predicted as NCR structures with CLscores below 0.5.

## S5.4 Analysis of Different Type of Errors by MOFClassifier

**Table S3**. Different types of NCR structures successfully identified by MOFClassifier. "Total Structurers" of each error type is flagged by Chen-Manz or MOFChecker when the structure flagged as NCR by all 3 checkers."

| Error Type | Total Structures | Flagged by MOFClassifier | Ratio |
|---|---|---|---|
| Overlapping Atoms | 5,916 | 5,865 | 99.14% |
| Isolated | 1,928 | 1,900 | 98.55% |
| Over-bond | 6,808 | 6,762 | 99.32% |
| Under-bond | 3,852 | 3,742 | 97.14% |
| Other | 831 | 816 | 98.19% |

# S6 Other Database

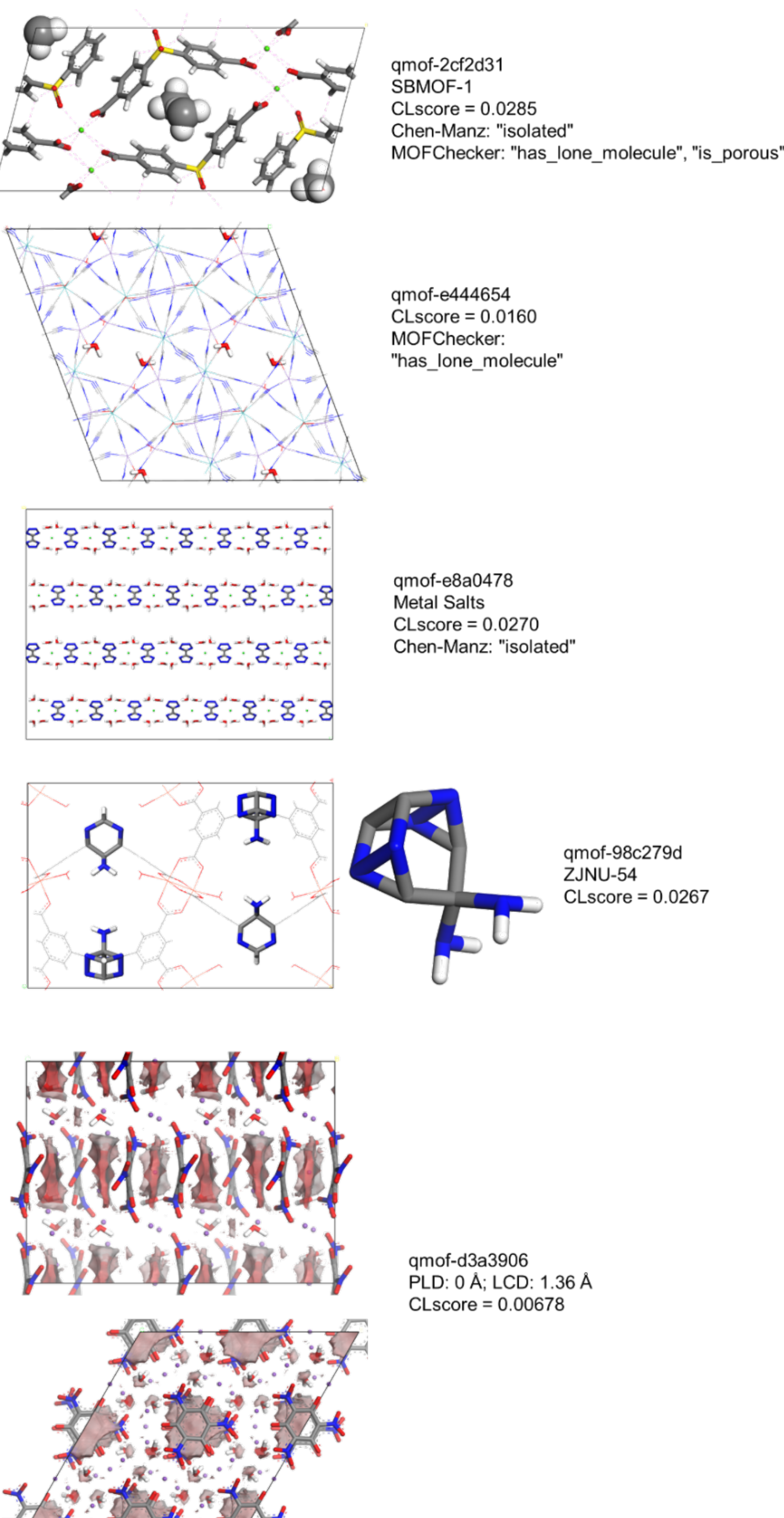

**Figure S25**. Representative NCR cases in the QMOF database are identified by the low CLscores predicted by MOFClassifier. Note that for qmof-d3a3906 the pore channel is shown instead of the framework.

## S7 Fine-tuned MOFClassifier

Considering that a small PLD does not necessarily indicate chemical errors in the crystal structure but rather serves as a criterion for identifying MOFs (as applied in the CoRE MOF database, CSD database, and MOFChecker), we integrated the QMOF dataset with the original data and retrained the MOFClassifier-QSP (with QMOF, including structures with small PLD). The training settings remained consistent with the previous configuration, and the dataset was expanded to 58,883 points (27,582 positive and 31,301 unlabeled).

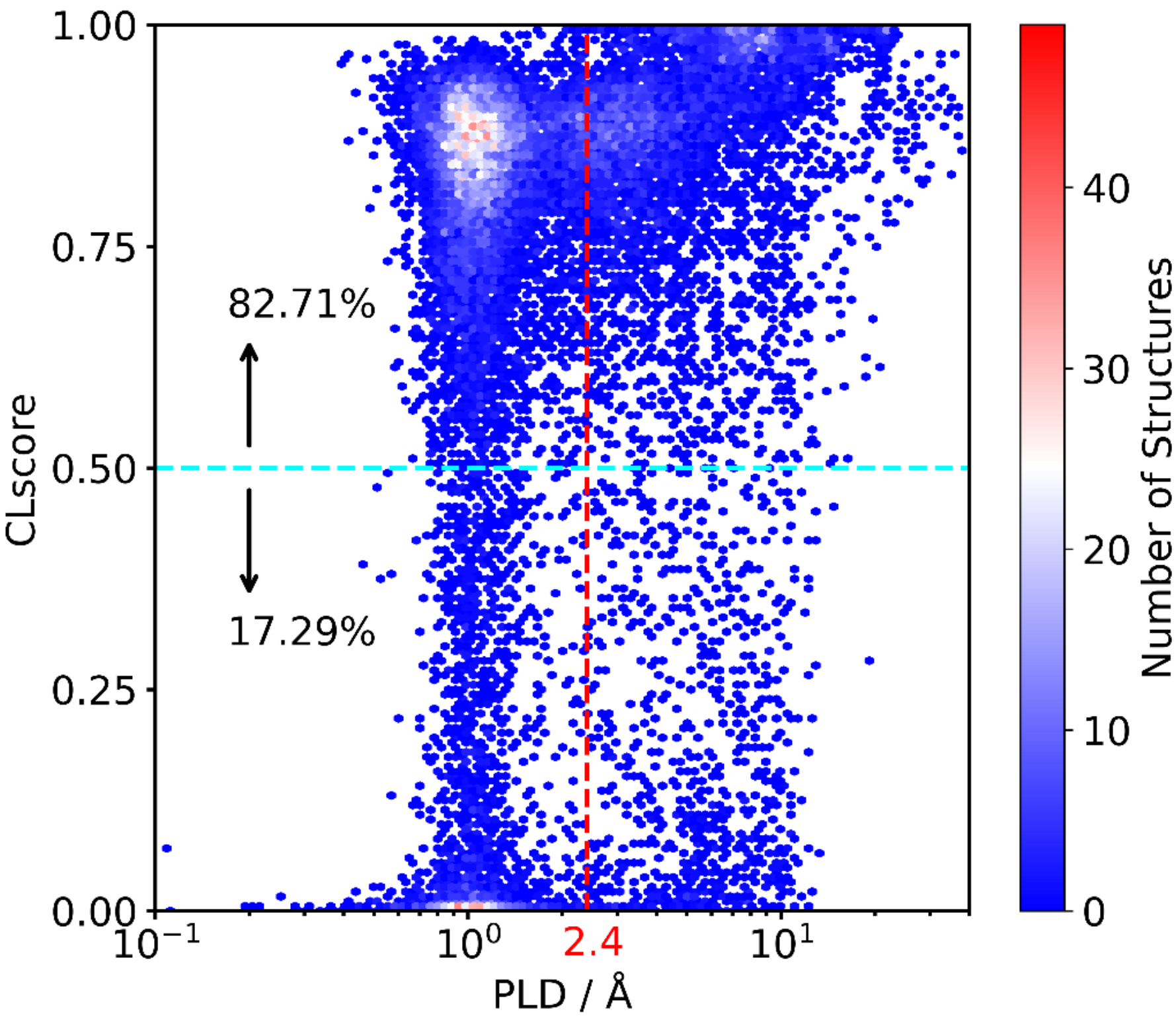

**Figure S26**. Scatter plot of PLD versus the CLscore predicted by MOFClassifier for QMOF DB. The color of each point represents point density. The red dashed line indicates the cutoff value of PLD = 2.4 Å, used to determine whether a structure is considered porous. The cyan dashed line represents the cutoff value distinguishing CR and NCR structures.

Based on the above results, MOFClassifier can be adapted to different MOF databases through fine-tuning. Additionally, once the dataset is updated and refined-through molecular simulations or manual inspection-we will retrain the model to further improve its accuracy. Finally, we outlook developing a transformer-based MOFClassifier to achieve model interpretability, enabling us to assess whether the model can accurately identify the exact locations of structural errors.

## S8 Benchmark for Run Time

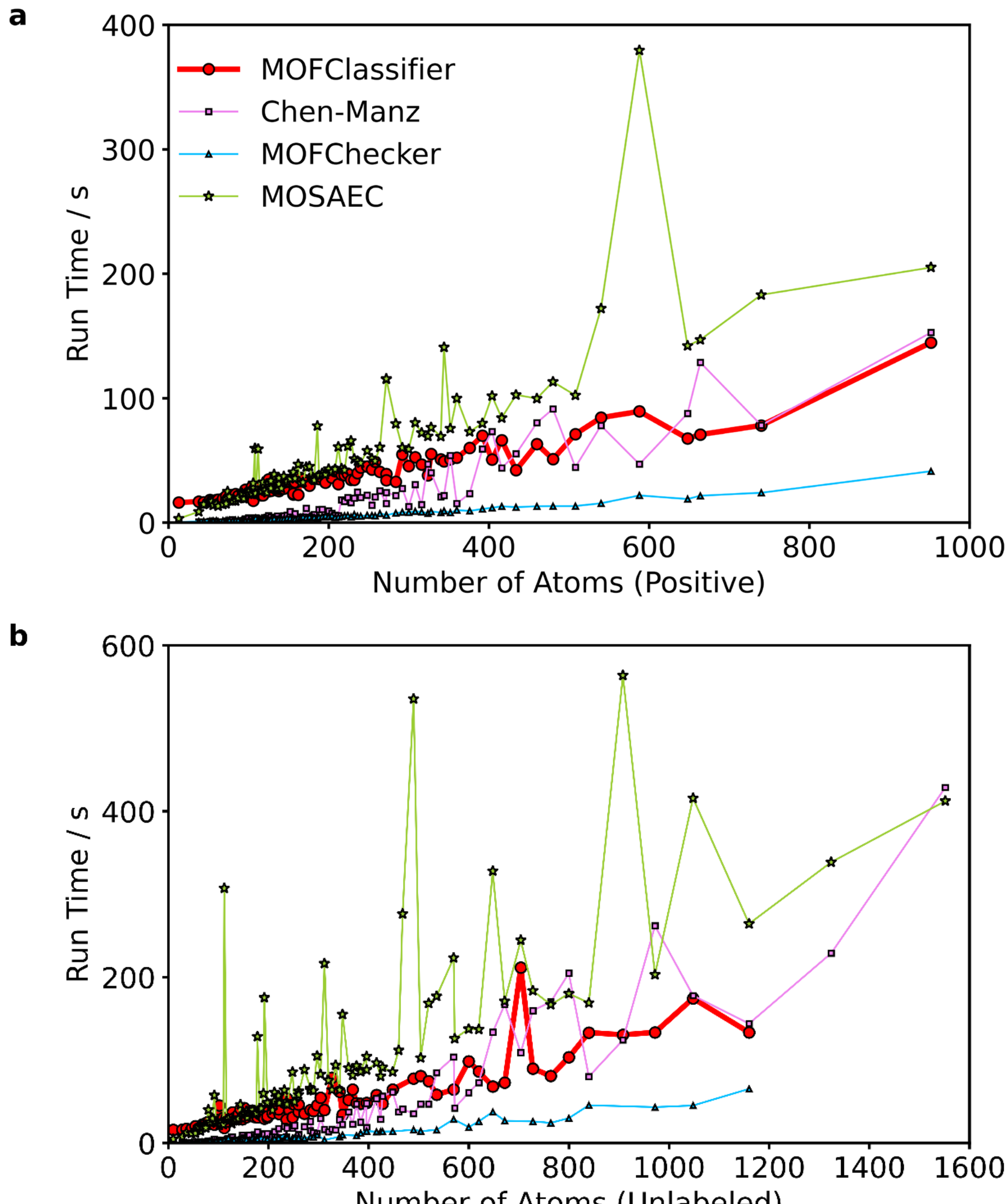

**Figure S27**. The run time of different methods for validation of MOFs from (**a**) positive and (**b**) unlabeled dataset. Note that we run the code for structures one by one instead of by batch. Some points are missing in the unlabeled samples of MOFClassifier and MOFChecker because the structures have severe disorder to be parsed by Pymatgen, making it impossible to generate crystal graphs and proceed with subsequent calculations.

**Table S4**. The prediction time for 10,000 randomly selected MOFs (3,482 positive and 6,518 unlabeled) using different batch sizes. The tests were conducted on an Intel Xeon Gold 6426Y CPU with 64 cores and a single NVIDIA GeForce RTX 4090 GPU.

| **Batch Size** | **Run Time (CPUs) / h** | **Run Time (GPU) / h** |
|---|---|---|
| 16 | 1.13 | 0.73 |
| 64 | 1.55 | 0.75 |
| 128 | 1.44 | 0.75 |

# S9 Impact on High-throughput Screening Results

**Table S5**. The CLscore < 0.5 (NCR) structures of top candidates (Total) from high-throughput screening or machine learning prediction according to other literature. For the CLscore of the structures from QMOF database, we used MOFClassifier-QSP model.

| Database | NCR / Total | Purpose | Reference |
|---|---|---|---|
| CoRE MOF 2019 | 7 / 15 | $CF_4/N_2$ | *J. Cleaner Prod.,* ***2024****, 461: 142634.* |
| CoRE MOF 2019 | 5 / 10 | $CO_2$ Separation | *Ind. Eng. Chem.,* ***2024****, 63(38): 16497-16508.* |
| CoRE MOF 2019 | 3 / 9 | thiophene/benzene | *J. Environ. Chem. Eng.,* ***2025****, 13(2): 116078.* |
| CoRE MOF 2019 | 36 / 42 | remove heavy metals | *Sep. Purif. Technol.,* ***2024****, 339: 126732.* |
| CoRE MOF 2019 | 6 / 15 | $NF_3/N_2$ | *Sep. Purif. Technol.,* ***2025****, 364: 132481.* |
| QMOF | 2 / 14 | $CO_2$ reduction | *Mater. Horiz.,* ***2024****,11: 4311-4320* |
| QMOF | 2 / 40 | $C_3H_8/N_2/O_2/Ar$<br>$C_3H_8/C_2H_6/CH_4$ | *J. Chem. Phys.,* ***2024****, 160 (8).* |
| CoRE MOF 2019 | 17 / 46 | $CO_2/CH_4/N_2/H_2$ | *Ind. Eng. Chem.,* ***2025*** *64 (5): 2926-2936.* |

We manually validated all structures with a CLscore < 0.5, among which 17 structures (out of 78) were validated as "good" through manual inspection. Of these 17 structures, only two were not flagged by any of the three checkers. Among the 17 manually validated "good" structures, Chen-Manz classified 9 as NCR, MOFChecker flagged 5, and MOSAEC flagged 5. Note that there may be overlaps among these checkers. This further confirms the influence of the dataset on model performance, suggesting that a more refined dataset could be used in the future to fine-tune the model.

## References

(1) Coudert, F.-X. Using Social Media Bots to Keep up with a Vibrant Research Field: The Example of @ MOF_papers. *Chem. Mater.* **2023**, *35* (7), 2657-2660. DOI: 10.1021/acs.chemmater.3c00383.

(2) Chen, T.; Manz, T. A. Identifying misbonded atoms in the 2019 CoRE metal-organic framework database. *RSC Adv.* **2020**, *10* (45), 26944-26951. DOI: 10.1039/D0RA02498H.

(3) Jin, X.; Jablonka, K. M.; Moubarak, E.; Li, Y.; Smit, B. MOFChecker: A Package for Validating and Correcting Metal-Organic Framework (MOF) Structures. *Digital Discovery* **2025**. DOI: 10.1039/D5DD00109A.

(4) White, A. J.; Gibaldi, M.; Burner, J.; Mayo, R. A.; Woo, T. K. High Structural Error Rates in "Computation-Ready" MOF Databases Discovered by Checking Metal Oxidation States. *J. Am. Chem. Soc.* **2025**. DOI: 10.1021/jacs.5c04914.

(5) Zhao, G.; Brabson, L. M.; Chheda, S.; Huang, J.; Kim, H.; Liu, K.; Mochida, K.; Pham, T. D.; Terrones, G. G.; Yoon, S. CoRE MOF DB: A curated experimental metal-organic framework database with machine-learned properties for integrated material-process screening. *Matter* **2025**. DOI: 10.1016/j.matt.2025.102140.

(6) Ouellette, W.; Darling, K.; Prosvirin, A.; Whitenack, K.; Dunbar, K. R.; Zubieta, J. Syntheses, structural characterization and properties of transition metal complexes of 5,5'-(1,4-phenylene) bis (1H-tetrazole)($H_2$bdt), 5',5''-(1,1'-biphenyl)-4,4'-diylbis (1H-tetrazole)($H_2$dbdt) and 5,5',5''-(1,3,5-phenylene) tris (1H-tetrazole)($H_3$btt). *Dalton Trans.* **2011**, *40* (45), 12288-12300. DOI: 10.1039/C1DT11590A.

(7) Zhao, G.; Chung, Y. G. PACMAN: A Robust Partial Atomic Charge Predicter for Nanoporous Materials Based on Crystal Graph Convolution Networks. *J Chem Theory Comput* **2024**, *20* (12), 5368-5380. DOI: 10.1021/acs.jctc.4c00434.

(8) Rosen, A. S.; Iyer, S. M.; Ray, D.; Yao, Z.; Aspuru-Guzik, A.; Gagliardi, L.; Notestein, J. M.; Snurr, R. Q. Machine learning the quantum-chemical properties of metal-organic frameworks for accelerated materials discovery. *Matter* **2021**, *4* (5), 1578-1597. DOI: 10.1016/j.matt.2021.02.015.

(9) Kim, W.-T.; Lee, W.-G.; An, H.-E.; Furukawa, H.; Jeong, W.; Kim, S.-C.; Long, J. R.; Jeong, S.; Lee, J.-H. Machine learning-assisted design of metal–organic frameworks for hydrogen storage: A high-throughput screening and experimental approach. *Chem. Eng. J.* **2025**, *507*, 160766. DOI: 10.1016/j.cej.2025.160766.

(10) Wang, R.; Zou, Y.; Zhang, C.; Wang, X.; Yang, M.; Xu, D. Combining crystal graphs and domain knowledge in machine learning to predict metal-organic frameworks performance in methane adsorption. *Microporous Mesoporous Mater.* **2022**, *331*, 111666. DOI: 10.1016/j.micromeso.2021.111666.

(11) Kang, Y.; Park, H.; Smit, B.; Kim, J. A multi-modal pre-training transformer for universal transfer learning in metal–organic frameworks. *Nat. Mach. Intell.* **2023**, *5* (3), 309-318. DOI: 10.1038/s42256-023-00628-2.

(12) Khan, S. T.; Moosavi, S. M. Connecting metal-organic framework synthesis to applications using multimodal machine learning. *Nat. Commun.* **2025**, *16* (1), 5642. DOI: 10.1038/s41467-025-60796-0.

(13) Mordelet, F.; Vert, J.-P. A bagging SVM to learn from positive and unlabeled examples. *Pattern Recognition Letters* **2014**, *37*, 201-209. DOI: 10.1016/j.patrec.2013.06.010.